\documentclass[12pt]{iopart}
\pdfoutput=1
\usepackage{iopams}
\expandafter\let\csname equation*\endcsname\relax
\expandafter\let\csname endequation*\endcsname\relax
\usepackage{amsmath,amssymb,mathtools}
\usepackage{graphicx,color}
\usepackage{enumerate}

\usepackage[colorlinks,linkcolor=blue,urlcolor=blue,citecolor=blue]{hyperref}
 \newcommand{\erefss}[1]{\eqref{#1}}%
 \newcommand{\erefs}[1]{\eqref{#1}}%
 \newcommand{\frefss}[1]{figures~\ref{#1}} %
 \newcommand{\frefs}[1]{~\ref{#1}} %
 \newcommand{\ffref}[1]{Figure~\ref{#1}} %
 \newcommand{\ssref}[1]{Section~\ref{#1}}%
 \newcommand{\aref}[1]{\ref{#1}}%
  \newcommand{\arefs}[1]{Appendix}%

\newcommand\be{\begin{equation}} 
\newcommand\ee{\end{equation}} 
\begin{document}

\title{Exact distribution for work and stochastic efficiency of an isothermal machine}
\author{Deepak Gupta}  
\address{Raman Research Institute,  Bangalore - 560080, India}

\date{\today}

\begin{abstract}
We consider an isothermal machine composed of two Brownian particles (say particle A and B) connected by a harmonic spring. A constant load is attached to particle A, and the particle B is trapped in a harmonic confinement whose minimum is dragged with a constant velocity. Whole system is in contact with the heat bath of a constant temperature. We obtain the distribution of the work done on particle A and particle B, and transient fluctuation theorem for these quantities is tested in the weak coupling limit and for both small and large observation time. Moreover, we show that the transient fluctuation theorem for total work done on both particles is satisfied. Furthermore, we compute the stochastic efficiency which is the ratio of the work done against the load force on particle A and the work done on particle B of this machine. The probability density function for stochastic efficiency is computed for all time. Numerical simulations are also done to verify the analytical results. 
\end{abstract}


\maketitle

%%% To make the table of contents %%%%%
\noindent\rule{\hsize}{2pt}
\tableofcontents
\noindent\rule{\hsize}{2pt}
\markboth{Exact distribution for work and stochastic efficiency of an isothermal machine}{}
%%%%%%%%%%%%%%%%%%%%%%%%%%%%%%%%%%%%%%%

\section{Introduction}
The function of an engine is to convert one form of energy into another form. For example, heat engines converts the energy from the heat reservoir to a useful work, windmills converts the wind energy to the electrical energy as a wind turbine or to pump the water as a windpump, a turbine connected to an electric generator utilizes the energy from the flowing water to convert into an electrical energy, a refrigerator pumps the heat from the cold environment to the hot environment, etc. The performance of an engine depends upon the amount of output power it delivers in the expense of the input power. For instance, a heat engine \cite{Callen, Huang, Zemansky} extracts heat $Q_h$ from the hot reservoir at a temperature $T_h$ and dumps some amount of heat $Q_c$ in the cold reservoir at a temperature $T_c<T_h$ in a cyclic manner, and it generates useful work $W=Q_h-Q_c$. The efficiency $\eta$ of such engine is given by $\eta=W/Q_h$ and it is bounded above by the Carnot efficiency $\eta_c=1-T_c/T_h$, \emph{i.e,} $\eta\leq\eta_c$ where $\eta_c$ is maximum possible efficiency achieved by an engine operating  in a quasi-static limit and in a reversible fashion. Hence, an engine operating at Carnot efficiency has power (output work per unit time) zero and is practically useless to do a work in a reasonable amount of time.

In the macroscopic thermodynamics, the system is composed of large number of degrees of freedom. Consequently, fluctuations are negligible, and the observables such as heat, work, entropy change, etc., attain a definite value. In contrast, for a system having small number of degrees of freedom, these observables become stochastic quantities \cite{Sekimoto,Seifert-1,Seifert-2}. The probability density function of a stochastic observable contains much wide information than their ensemble average value. In past two and half decades, there have been lot of experimental \cite{Wang,Wang-2,Garnier2005,Douarche2005,Ciliberto-0,Ciliberto-1,Berut-2014}  and theoretical \cite{SS-1,SS-2,Apal-1,Apal-2,Verley-11,Mazonka1999,Vanzon-1,Vanzon-2,Vanzon-3,Visco,Kundu, Farago2002,Deepak, Trap, HT} investigations done to understand the probability density function and their fluctuation relations \cite{Evans-Cohen, Evans-Searles,Searles-Evans,Searles2001,Gallavotti-Cohen,Kurchan,Lebowitz-Spohn,Jarzynski-1,Jarzynski-2,Crooks1998,Crooks-1,JDnoh} associated with them.

For a microscopic machine or engine converting the input power into the output power, efficiency which is the ratio of the output power and the input power is a stochastic quantity. Recently, a number of studies have been devoted to investigate the probability density function and large deviation function \cite{Touchette} of the stochastic efficiency of a microscopic engine. For example, Varley et al. \cite{verley-1} considered two models of microscopic engine namely Brownian work-to-work converter and photoelectric device to study the large deviation of the stochastic efficiency. In a similar work, Verley et al. \cite{verley-2} obtained a method to compute the large deviation function for stochastic efficiency from the characteristic function of input and output power. Subsequently, they used this method to study the efficiency statistics using a system consisted of two states coupled to two distinct temperature reservoirs. Gingrich et al. \cite{Gingrich} studied the efficiency fluctuation and the large deviation for a time-asymmetric stochastic heat engine consisted of a two states where the temperature and the energy levels are varied cyclically in four consecutive steps.  The probability density function and the large deviation function for the stochastic efficiency of effusion as a thermal engine were investigated by Proesmans et al. \cite{Proesmans-1}. Polettini et al. \cite{Polettini} derived the full probability density function for the stochastic efficiency when the thermodynamics fluxes obeys the multivariate Gaussian distribution with cumulants proportional to the time of observation. Moreover, it was shown that the  probability density function for efficiency  has two maxima  and one minimum. Proesmans et al. \cite{Proesmans-3} studied an isothermal engine using a Brownian particle driven by two time-periodic external forces where one force serves as load and the other one plays the role of a drive. The statistics of the stochastic efficiency is obtained analytically and also verified with the experiment. In Ref. \cite{Deepak-eff}, authors generalized a model of isothermal work-to-work converter engine given in Ref. \cite{verley-1} in the underdamped limit using the stochastic external load and stochastic drive force instead of constant external forces. They obtained the large deviation function and large but finite time probability density function for stochastic efficiency of an isothermal engine. Some more studies in this area one can see in the Refs. \cite{Proesmans-2,Proesmans-4,Benenti, Naoto, Park}. Moreover, several experiments have also been performed to understand the efficiency of a microscopic engine  \cite{Blickle,Martinez,Sudeesh}.

In the paper, we consider a one dimensional isothermal engine composed of two Brownian particle (say A and B) interacting with each other with harmonic potential where particle B is confined in a harmonic trap. When time $t\leq0$, the minimum of the harmonic trap is stationary and is at the origin of $x$-axis. Hence, the system has equilibrium Boltzmann's distribution at $t=0$. At $t>0$, the minimum of the harmonic confinement is dragged with a constant velocity, and a load is attached to the Brownian particle A. We study the work done on particle A ($W_A$) and particle B ($W_B$), and the stochastic efficiency $\eta$ of the isothermal engine which is the ratio of the work done against the load on particle A to the work done on particle B by dragging the minimum of the harmonic potential with constant velocity. The probability density function for $W_A$, $W_B$ and $\eta$ are obtained for all time.

The paper is organized as follows. In \sref{model-sec}, we give the model system and definitions of work done on particle A, particle B, and the stochastic efficiency. \ssref{joint-sec} contains the joint distribution for the work done on particle A and particle B. The transient fluctuation theorem  \cite{Evans-Searles} is studied for work done on particle A and B in \sref{tft-1-2}. In \sref{tft}, we discuss the transient fluctuation theorem for total work done on the system. The probability density function for the stochastic efficiency is computed and phase diagrams are shown which suggests the sign of efficiency where the peak of the density function occurs in \sref{p-eta-sec}. We summarize our paper in \sref{summ-sec}. Some of the results are given in \arefs{app-sec}.

\section{Model}
\label{model-sec}
\begin{figure}
\begin{center}
    \includegraphics[width=10cm]{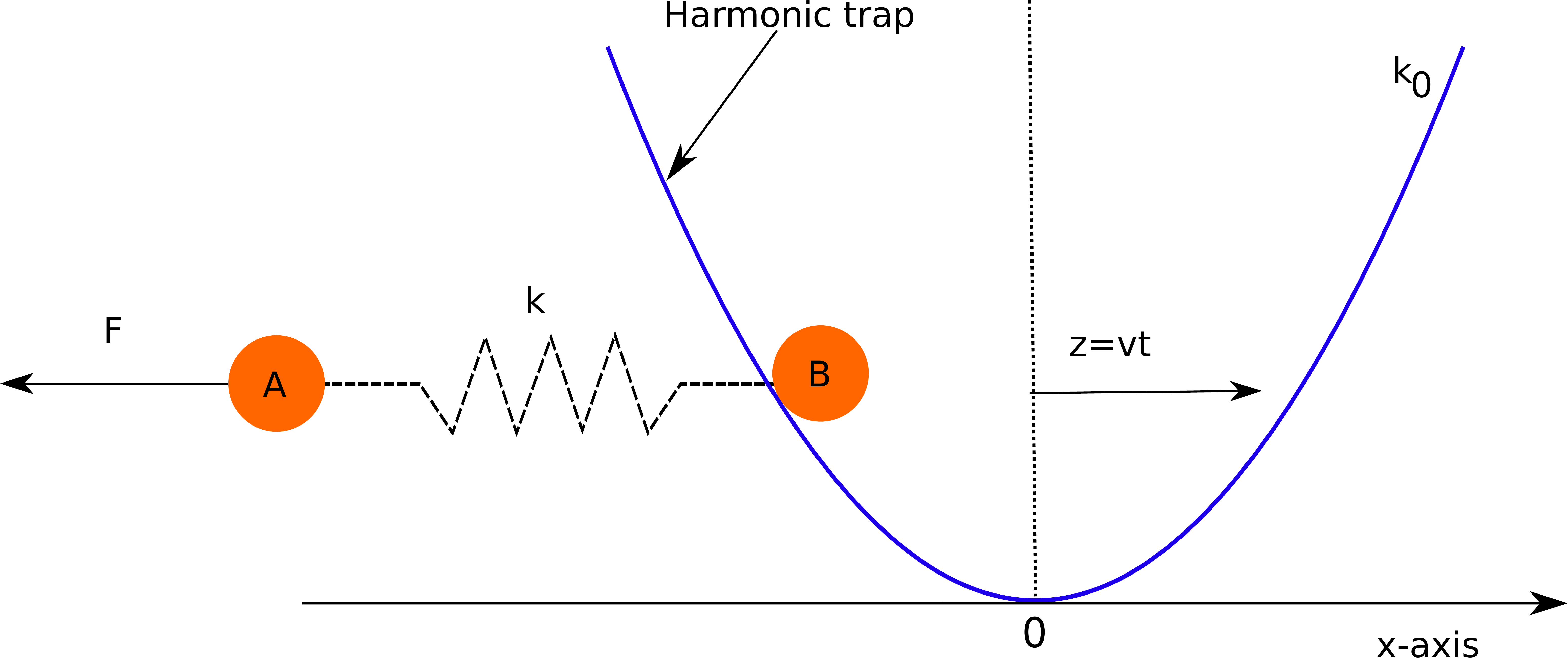}{\centering}
\end{center}
   \caption{\label{machine} The schematic diagram for an isothermal work-to-work converter machine is shown. The particle B is confined in a harmonic trap of stiffness constant $k_0$. The particle A is coupled harmonically with particle B with a spring of stiffness $k$. The whole setup is immersed in a heat bath (not shown) of constant temperature $T$. The minimum of the potential $z$ is moved with constant velocity $v$ from time $t>0$ to $t=\tau$, \emph{i.e.,} $z=vt$ for $t\in(0,\tau]$, and a load $F$ is attached to particle A. Sign of $F$ is taken to be negative if the load is pulling the particle A and sign of $v$ is taken to be positive if the minimum of the trap is moving towards positive direction of $x$-axis. The vertical dotted line indicates the location of minimum of the harmonic trap at time $t$.}
\end{figure}
Consider a model of a machine consists of two Brownian particles (say particle A and B) coupled by a harmonic spring of stiffness $k$. Suppose particle B is trapped in a harmonic confinement of stiffness $k_0$. The whole setup is immersed in a heat bath of a constant temperature $T$. The potential energy of the system is given by
\begin{equation}
V(x_A,x_B,t)=\dfrac{k}{2}(x_A-x_B)^2+\dfrac{k_0}{2}(x_B-z)^2,
\label{pot-eg}
\end{equation}
where $z$ is the minimum of the harmonic trap which is time dependent, and $x_A$ and $x_B$ are the positions of particle A and B, respectively.

Suppose the minimum of the harmonic trap is moved with a constant finite velocity $v$ and is at $z =vt$ at time $t>0$, and a constant finite load $F$ is attached to particle A from time $t>0$ to $t=\tau$. The schematic diagram of the machine is shown in \fref{machine}. Note that similar model for a Brownian particle confined in a harmonic potential ($k=0$) whose minimum is dragged with a given velocity is already studied both theoretically \cite{Mazonka1999,Vanzon-1,Vanzon-2,Vanzon-3} and experimentally \cite{Wang,Wang-2}.

The dynamics of the given system is described by following overdamped Langevin equations
\begin{align}
\gamma\dot{x}_A=&-k(x_A-x_B)+\xi_A(t)+F, \label{dyn-1}\\
\gamma\dot{x}_B=&-k(x_B-x_A)+\xi_B(t)-k_0(x_B-z), \label{dyn-2}
\end{align}
where dot represents the derivative with respect to time, $\gamma$ is the dissipation constant, $\xi_A(t)$ and $\xi_B(t)$ are the thermal noises acting on particle A and B, respectively, from the heat bath, having mean zero and correlations $\langle\xi_i(t)\xi_j(t^\prime)\rangle=2 \gamma T\delta_{ij}\delta(t-t^\prime)$. We set Boltzmann's constant $k_B=1$ throughout the calculation.
 
Multiplying \eref{dyn-1} by $\dot{x}_A(t)$ and \eref{dyn-2} by $\dot{x}_B(t)$ on both sides, integrating over time from $t=0$ to $t=\tau$, and adding them together, yields first law of thermodynamics,
\begin{equation}
\Delta V=W_A+W_B+Q_A+Q_B,
\end{equation} 
in which 
\begin{align}
\Delta V=&\dfrac{1}{T}[V(x_A(\tau),x_B(\tau),\tau)-V(x_A(0),x_B(0),0)],\\
W_A=&\dfrac{F}{T}\int_0^\tau dt\ \dot{x}_A(t)=\dfrac{F}{T}[x_A(\tau)-x_A(0)],\label{WA}\\
W_B=&\dfrac{v\gamma}{T\tau_\gamma}\int_0^\tau dt\ [z-x_B(t)],\label{WB}\\
Q_A=&\dfrac{1}{T}\int_0^\tau dt\ [\xi_A(t)-\gamma \dot{x}_A(t)]\dot{x}_A(t), \label{QA}\\
Q_B=&\dfrac{1}{T}\int_0^\tau dt\ [\xi_B(t)-\gamma \dot{x}_B(t)]\dot{x}_B(t). \label{QB}
\end{align}
In \eref{WB}, $\tau_\gamma=\gamma/k_0$ is the characteristic time scale. Here, the observables change in the potential energy $\Delta V$, work done ($W_A$ and $W_B$) on particle A and B, and heat absorbed ($Q_A$ and $Q_B$) by particle A and B are measured in the unit of temperature $T$ of the heat bath and with respect to the initial steady state distribution given in \eref{IC}. The integrals shown in \eref{QA} and \eref{QB}  follow the Stratonovich rule of integration.

The observable in this paper is efficiency $\eta$ of the machine which is the ratio of work done ($-W_A$) against the load force $F$ on particle A to the work done ($W_B$) on  particle B from time $t>0$ to $t=\tau$, \emph{i.e.,}
\begin{equation}
\eta=-\dfrac{W_A}{W_B}.
\end{equation}
In the above expression, both $W_A$ and $W_B$ are stochastic quantities. Therefore, the efficiency is also a stochastic observable. The probability density function for the stochastic efficiency $p_\tau(\eta)$ is computed as follows:
\begin{align}
p_\tau(\eta)=&\int_{-\infty}^{+\infty}dW_A \int_{-\infty}^{+\infty}dW_B\ P_\tau(W_A,W_B)\ \delta(\eta+W_A/W_B)\nonumber\\
=&\int_{-\infty}^{+\infty}dW_B\ |W_B|\ P_\tau(-\eta W_B, W_B),\label{p-eta-int}
\end{align} 
where $P_\tau(W_A,W_B)$ is the joint distribution of $W_A$ and $W_B$ at time $\tau$, and $|W_B|$ is the Jacobian.

%%%%%%%%%%%%%%%%%%%%%%%%%%%%%%%%%%%%%%%%%%%%%%%%%%%%%%%%%%%%%%%%%
\section{Joint distribution $P_\tau(W_A,W_B)$}
\label{joint-sec}
In this paper, our aim is to compute the efficiency of a machine which does work against the load $F$ attached to particle A by dragging the harmonic trap which confines the particle B from time $t>0$ to $t=\tau$. When time $t\leq0$, there was no load attached to particle A, and the minimum of the trap was at the origin, \emph{i.e.,} $z=0$. Thus, the system obeys the steady state distribution at time $t=0$:
\begin{equation}
P(U_0)=\dfrac{1}{\sqrt{(2\pi)^2 \det{\Sigma}}}\exp\bigg[-\dfrac{1}{2}U_0^T \Sigma^{-1} U_0\bigg],\label{IC}
\end{equation}
where the row vector $U_0^T=[x_A(0),x_B(0)]$, and the correlation matrix 
\begin{equation}
\Sigma=\dfrac{T\tau_\gamma}{\gamma}\begin{pmatrix}
1+1/\delta&&1\\
1&&1
\end{pmatrix},
\label{sigma}
\end{equation}
where $\delta=k/k_0$ is the dimensionless coupling parameter.

For $t>0$, the dynamics of the system given in \eref{dyn-1} and \eref{dyn-2}, can be rewritten in a matrix form as
\begin{equation}
\dfrac{dU}{dt}=-\dfrac{1}{\tau_\gamma}AU+\dfrac{1}{\gamma}\zeta(t)+B(t),
\end{equation}
where column vectors $\zeta(t)=[\xi_A(t),\xi_B(t)]^T$, $B(t)=(F/\gamma,z/\tau_\gamma)^T$, and the matrix $A$ is
\begin{equation*}
A=\begin{pmatrix}
\delta&&-\delta\\
-\delta&&1+\delta
\end{pmatrix}.
\end{equation*}
The solution of above equation at time $0<t\leq\tau$ is given by
\begin{equation}
U(t)=G(t)U_0+\dfrac{1}{\gamma}\int_0^t dt^\prime G(t-t^\prime)[\gamma B(t^\prime)+\zeta(t^\prime)],
\end{equation}
where the symmetric matrix $G(t)=e^{-(t/\tau_\gamma)A}$ is given in \aref{Green-mat}.

Therefore, one can obtain the mean and correlation of $U(t)$ as
\begin{align}
\overline{\langle U(t)\rangle}&=\int_0^\tau dt^\prime G(t-t^\prime) B(t^\prime),\label{mean-U}\\
\overline{\langle M(t)M^T(t)\rangle}&=\Sigma,\label{var-U}
\end{align}
where superscript $T$ refers the transpose of a matrix, $\Sigma$ is given in \eref{sigma}, and $M(t)= U(t)-\overline{\langle U(t)\rangle}$. The explicit form of mean of $U(t)$ is given in \aref{mean-U-sec}. In the above equations, the angular brackets and overhead bar represent the averaging over noises and the initial state $U_0$ with respect to steady state distribution $P(U_0)$ [see \eref{IC}], respectively.

The work done on the particle A and B at time $\tau$ are given in \eref{WA} and \eref{WB}, respectively. Both of these quantities are linear in thermal noises. Therefore, it is sufficient to compute the means and correlations of them to write the joint distribution $P_\tau(W_A,W_B)$, and these are given by
\begin{align}
\mu_A&=\dfrac{F}{T}\overline{\langle x_A(\tau)\rangle},\label{mean-A}\\
\mu_B&=\dfrac{v^2\tau^2\gamma}{2 T\tau_\gamma}-\dfrac{v\gamma}{T\tau_\gamma}\int_0^\tau  dt\ \overline{\langle x_B(t) \rangle},\label{mean-B}\\
C_{AA}&=\dfrac{2F^2}{T^2}\big[\{1-G_{11}(\tau)\}\overline{x_A^2(0)}-G_{12}(\tau)\overline{x_A(0)x_B(0)}\big],\label{var-A}\\
C_{BB}&=\dfrac{v^2\gamma^2}{T^2\tau_\gamma^2}\int_0^\tau dt_1\ \int_0^\tau dt_2\ \overline{\langle M_{21}(t_1)M_{21}(t_2) \rangle},\label{var-B}\\
C_{AB}&=-\dfrac{Fv\gamma}{T^2\tau_\gamma}\int_0^\tau dt\ \overline{\langle M_{21}(t)[M_{11}(\tau)-x_A(0)]\rangle}=0\label{corr-AB},
\end{align}
where the matrix elements $G_{ij}(t)=[G(t)]_{ij}$, $M_{i1}(t)=[M(t)]_{i1}$, and $\mu_r=\overline{\langle W_r\rangle}$, $C_{rl}=\overline{\langle [W_r-\overline{\langle W_r\rangle}][W_l-\overline{\langle W_l\rangle}]\rangle}$ with $\{i,j\}=\{1,2\}$ and $\{r,l\}=\{A,B\}$. In \eref{var-A}, $\overline{x_A^2(0)}=[\Sigma]_{11}$ and $\overline{x_A(0)x_B(0)}=[\Sigma]_{12}$. The explicit form of these means and correlations are given in \aref{mean-corr} in which the dimensionless parameters $\alpha=F/\sqrt{\gamma T}$ and $\theta=2v\gamma/F$ are the strength of force acting on particle A with respect to that of bath and relative strength acting on harmonic trap which confines particle B to that on particle A, respectively.   

Since the correlation between $W_A$ and $W_B$ is zero: $C_{AB}=0$, the joint distribution $P_\tau(W_A,W_B)$ can be written in factorize form: $P_\tau(W_A,W_B)=P_A(W_A)P_B(W_B)$, where 
\begin{equation}
P_r(W_r)=\dfrac{1}{\sqrt{2 \pi C_{rr}}} \exp\bigg[-\dfrac{(W_r-\mu_r)^2}{2 C_{rr}}\bigg],    \quad\quad r=A,B.
\label{joint}
\end{equation}
For convenience, we have dropped the subscript $\tau$ in $P_r(W_r)$.
\begin{figure*}
 \begin{tabular}{cc}
    \includegraphics[width=7cm]{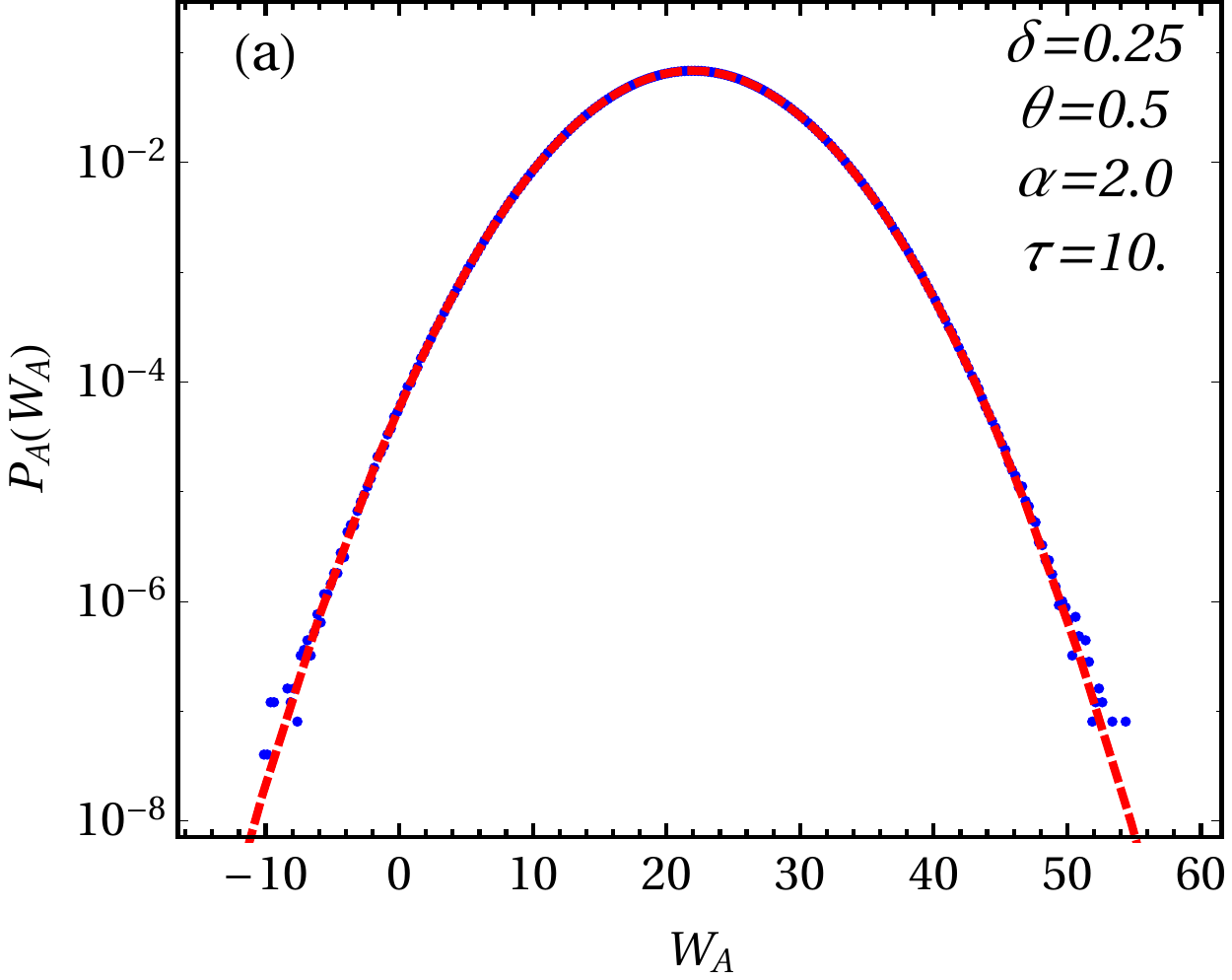}
    \includegraphics[width=7cm]{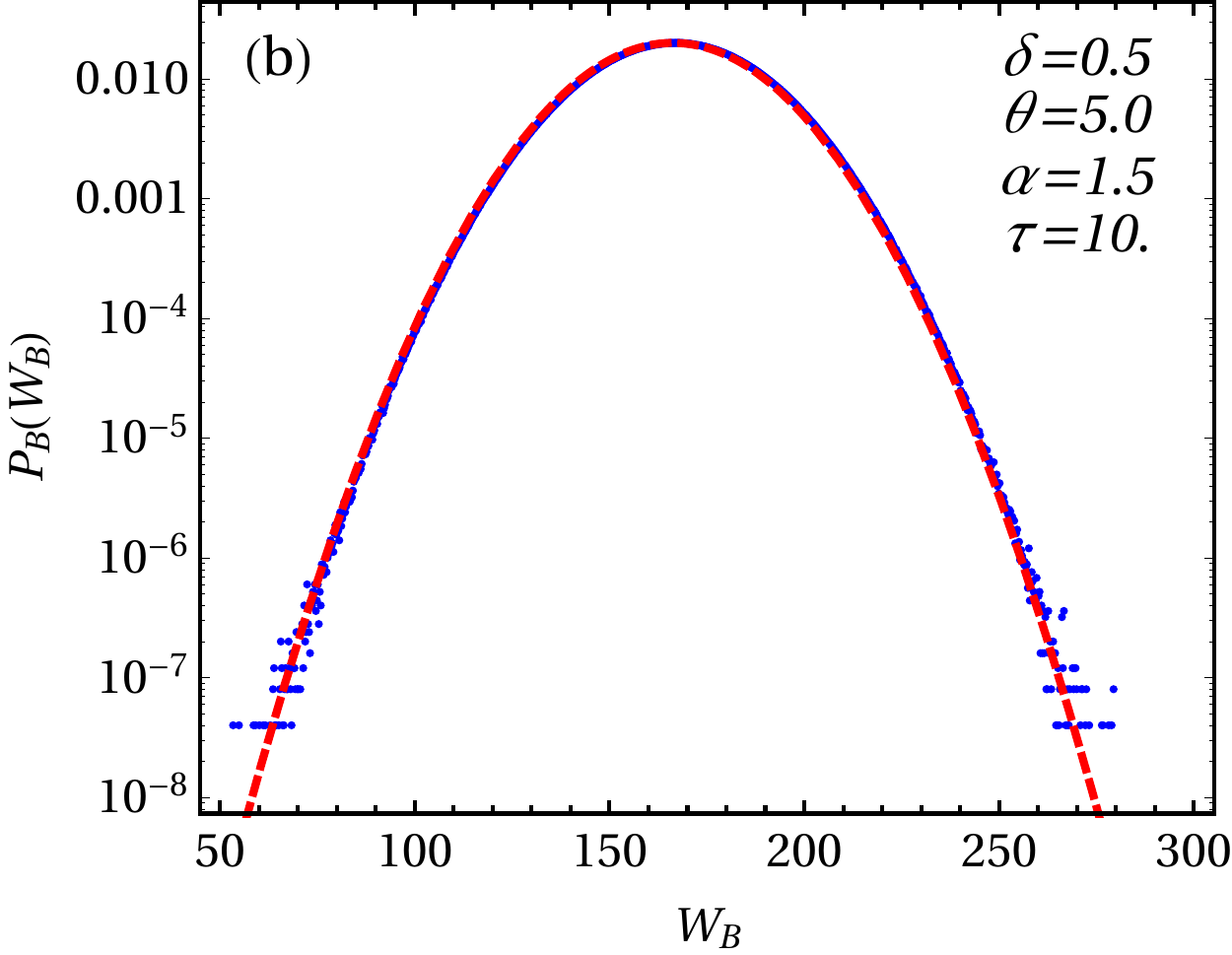}
    \end{tabular}
 \bigskip
   \caption{\label{pdf-prob-W1-W2}The probability density function for work done on particle A and B are shown for: (a) for  $\theta=0.5$, $\alpha=2.0$, coupling parameter $\delta=0.25$   at $\tau=10$, and (b) for $\theta=5.0$, $\alpha=1.5$, coupling parameter $\delta=0.5$ at time $\tau=10$. In both figures, red dashed lines are the analytical results given in \eref{joint}, and blue dots are obtained from the numerical simulations at the respective times $\tau$. These results are shown for fixed $\tau_\gamma=1$.}
\end{figure*}

In \fref{pdf-prob-W1-W2}, we have shown the comparison of the analytical result of the probability density function given in \eref{joint}, with the numerical simulation. The plots show that there is nice agreement between theory and numerical simulation.
%%%%%%%%%%%%%%%%%%%%%%%%%%%%%%%%%%%%%%%%%%%%%%%%%%%%%%%%%%%%%%%%%%%%%%%%%%%%%%%%%%%%%%%%%%%%%%%%%%%%%%%%%%%%%%%

\section{Transient fluctuation theorem for work done $W_A$ and $W_B$}
\label{tft-1-2}
For a system initially in the equilibrium and then driven away from the equilibrium using external driving, the stochastic quantity $\Omega$ is observed. Let $P(\Omega)$ be the probability density function of $\Omega$. When $\Omega$ satisfies the transient fluctuation theorem (TFT) \cite{Evans-Searles}, it obeys the following relation
\begin{equation}
\dfrac{P(\Omega)}{P(-\Omega)}=e^{\Omega}.
\end{equation}
The above relation states that the probability of getting positive values of $\Omega$ is exponentially favourable than that of negative values.

In our case, we analyze TFT for both $W_A$ and $W_B$. It is clear that TFT  may not hold for all parameters (see \aref{mean-corr}). However, we investigate TFT in the large and small time limit. 

In the weak coupling limit $(\delta\to0)$ and for small observation time ($u_\tau\ll\delta^{-1}$), the means and correlations given in \aref{mean-corr} reduce to
\begin{align}
\tilde{\mu}_A=&\alpha^2 u_\tau \tau_\gamma+O(\delta)\label{mA-wc-st},\\
\tilde{\mu}_B=&\dfrac{\alpha^2\theta^2\tau_\gamma}{4}(e^{-u_\tau}+u_\tau-1)+O(\delta),\label{mB-wc-st}\\
\tilde{C}_{AA}=&2\alpha^2 u_\tau \tau_\gamma+O(\delta)\label{cAA-wc-st},\\
\tilde{C}_{BB}=&\dfrac{\alpha^2\theta^2\tau_\gamma}{2}(e^{-u_\tau}+u_\tau-1)+O(\delta)\label{cBB-wc-st},
\end{align} 
where $u_\tau=\tau/\tau_\gamma.$

Using above equations, one can see that $\tilde{C}_{AA}=2 \tilde{\mu}_A$ and $\tilde{C}_{BB}=2 \tilde{\mu}_B$ in the limit $\delta\to 0$. Therefore, TFT for both work done ($W_A$ and $W_B$) is satisfied in the weak coupling limit and for small observation time ($u_\tau\ll\delta^{-1}$).
\begin{figure*}
 \begin{tabular}{cc}
    \includegraphics[width=7cm]{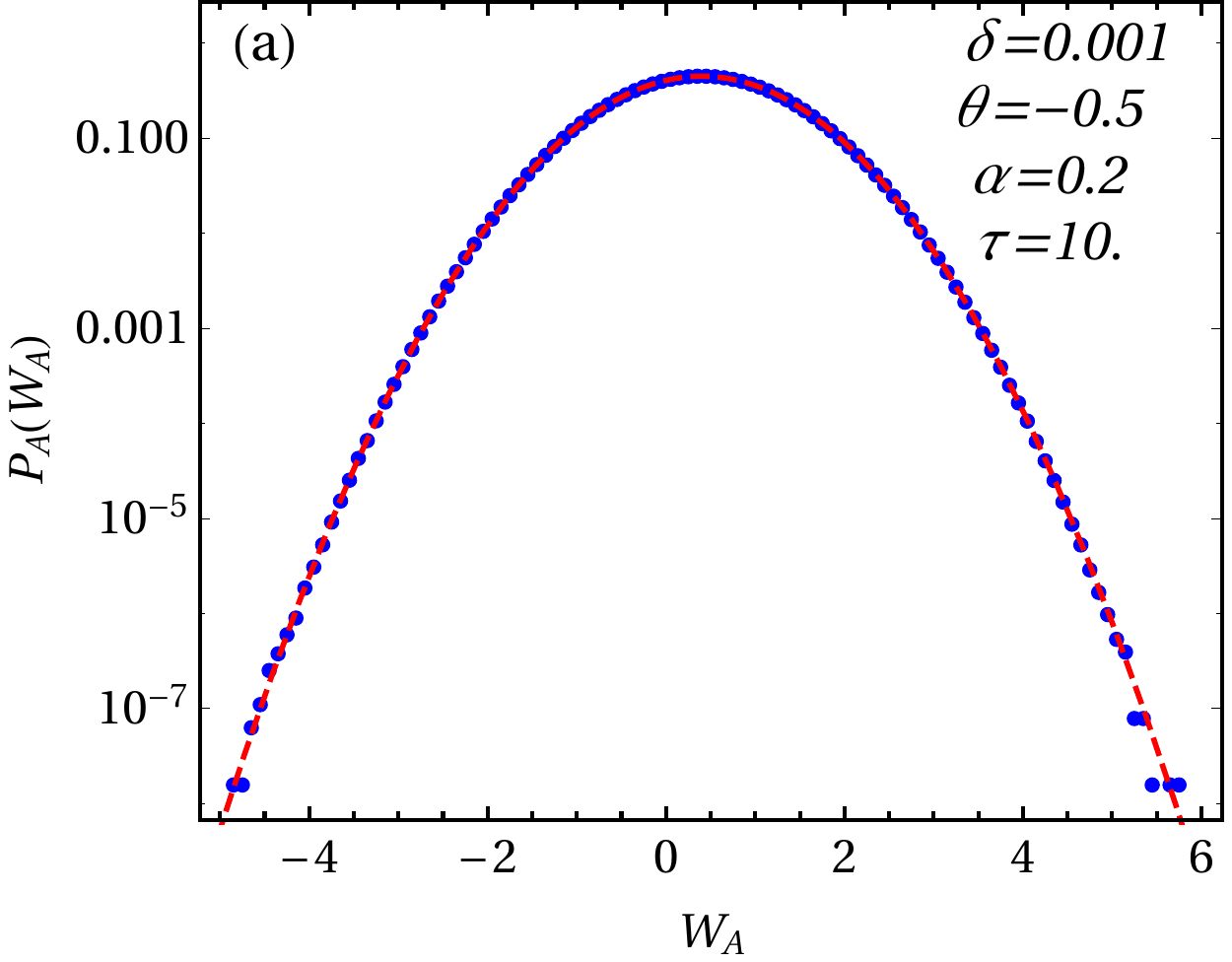}
    \includegraphics[width=6.7cm]{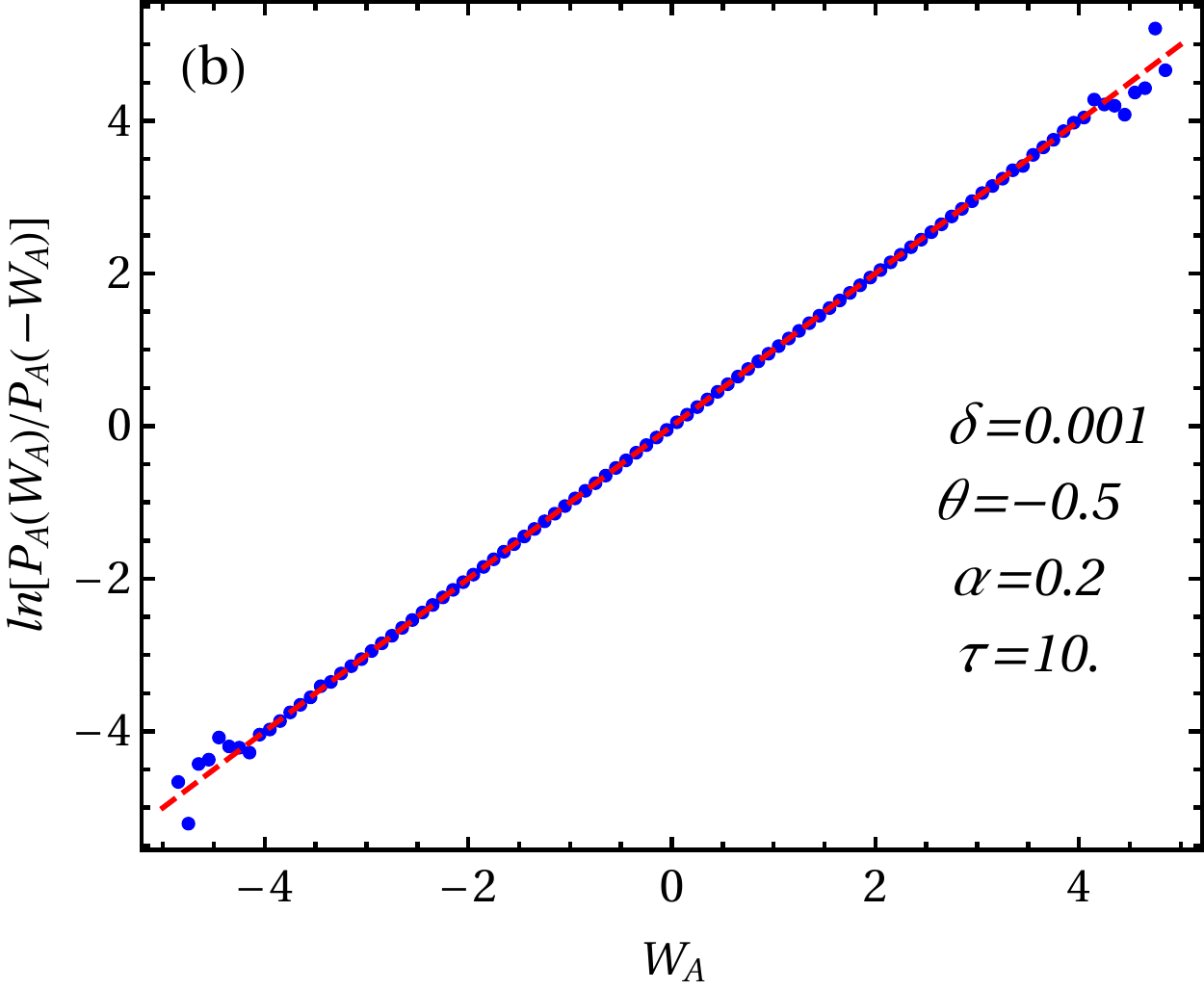}\\
\includegraphics[width=7cm]{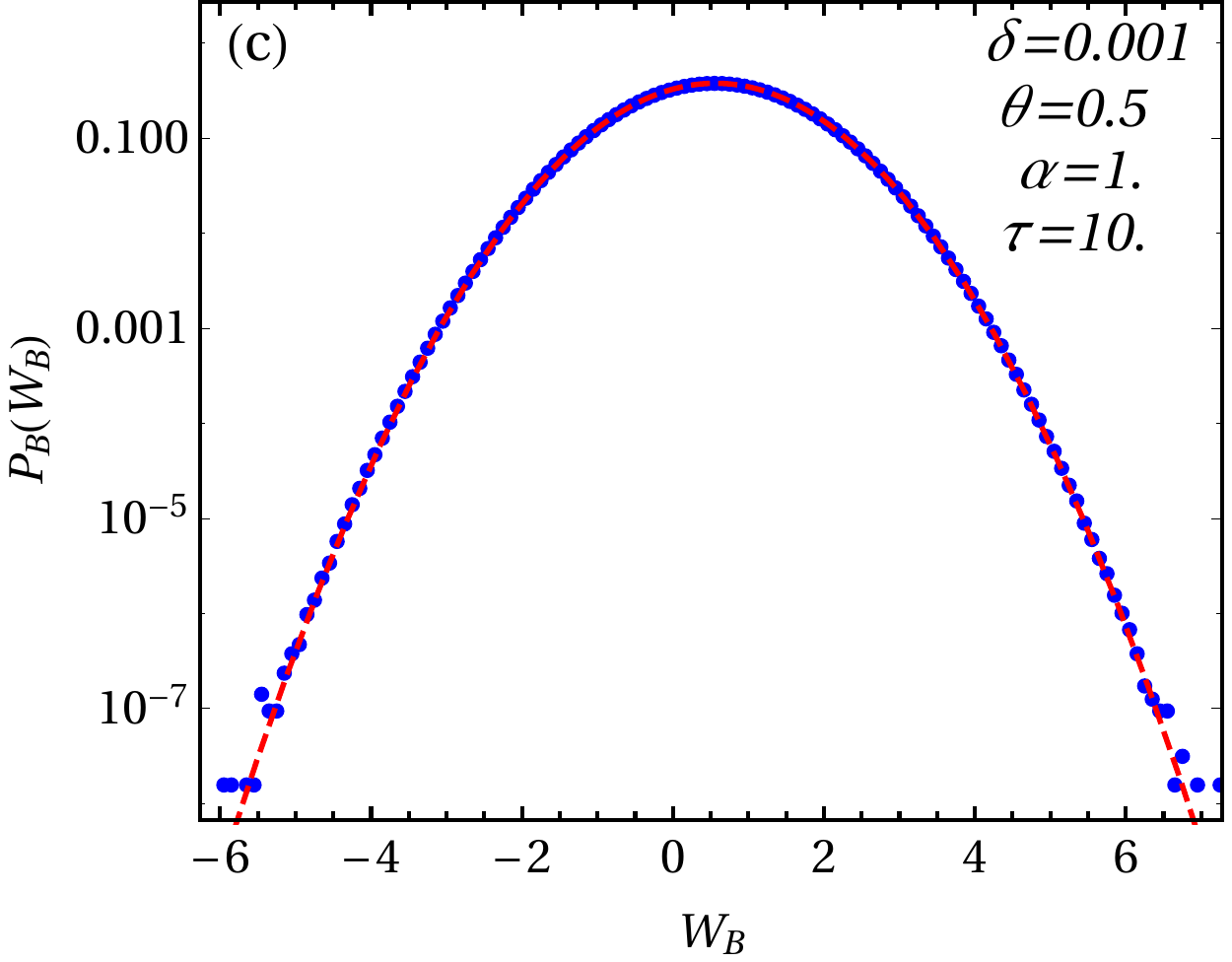}
    \includegraphics[width=6.7cm]{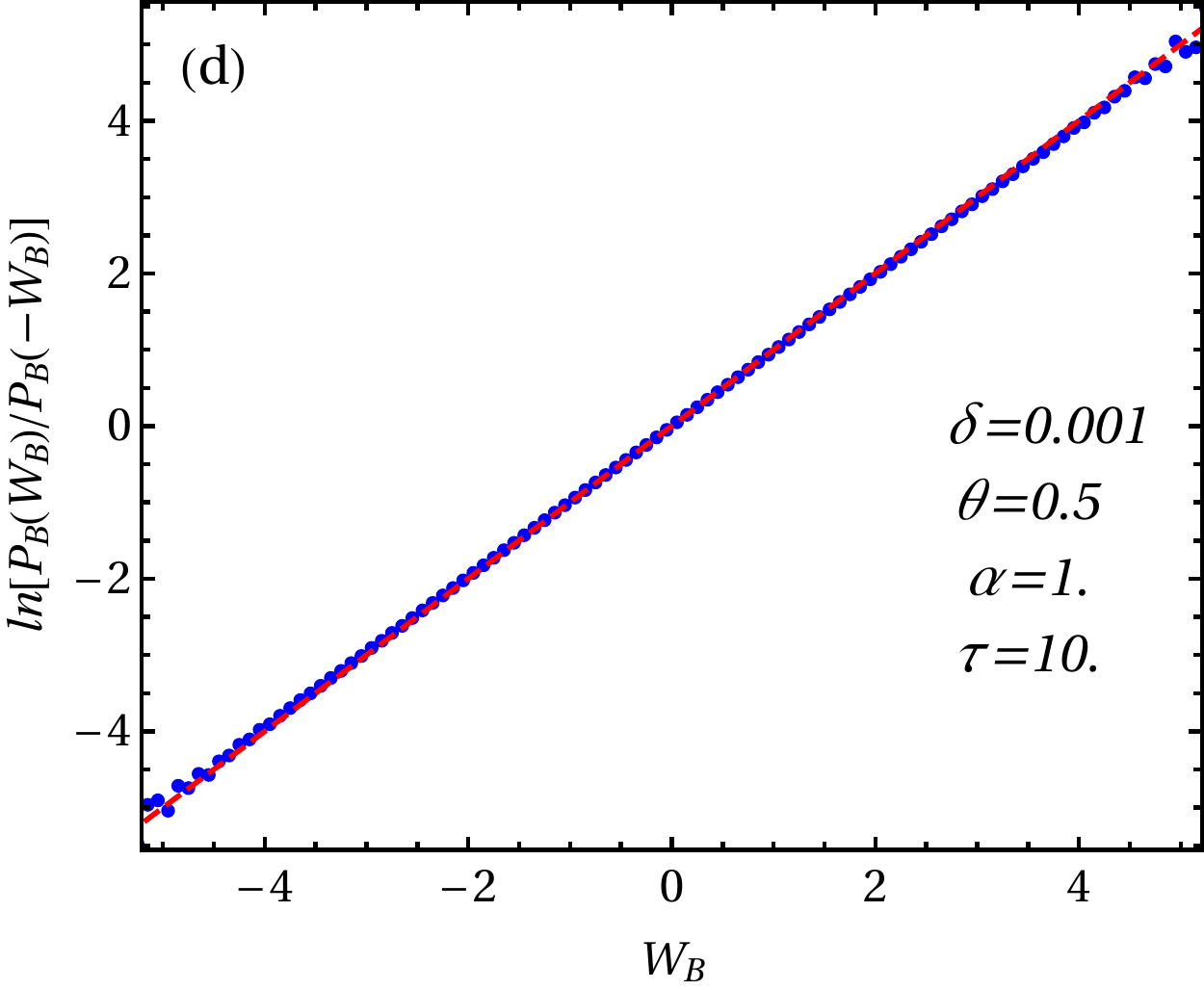}
    \end{tabular}
 \bigskip
   \caption{\label{small-delta-small-time} Figures (a) and (c) represent the probability density function for work done on particle A and B, respectively. The function $\ln[P_i(W_i)/P_i(-W_i)]$ are plotted against $W_i$ in (b) and (d). In all figures, the red dashed lines are the analytical result obtained from \eref{joint} with cumulants given in \erefss{mA-wc-st}--\erefs{cBB-wc-st} whereas the blue dots are obtained from numerical simulations. These results are shown for observation time $\tau\ll\delta^{-1}$ and for fixed $\tau_\gamma=1$. In figures (b) and (d), red dashed lines have slope unity which indicate that probability density functions for both work done ($W_A$ and $W_B$) satisfy TFT in the weak coupling limit $(\delta\ll1)$ and for small  observation time ($\tau\ll\delta^{-1}$). }
\end{figure*}
\begin{figure*}
 \begin{tabular}{cc}
    \includegraphics[width=7cm]{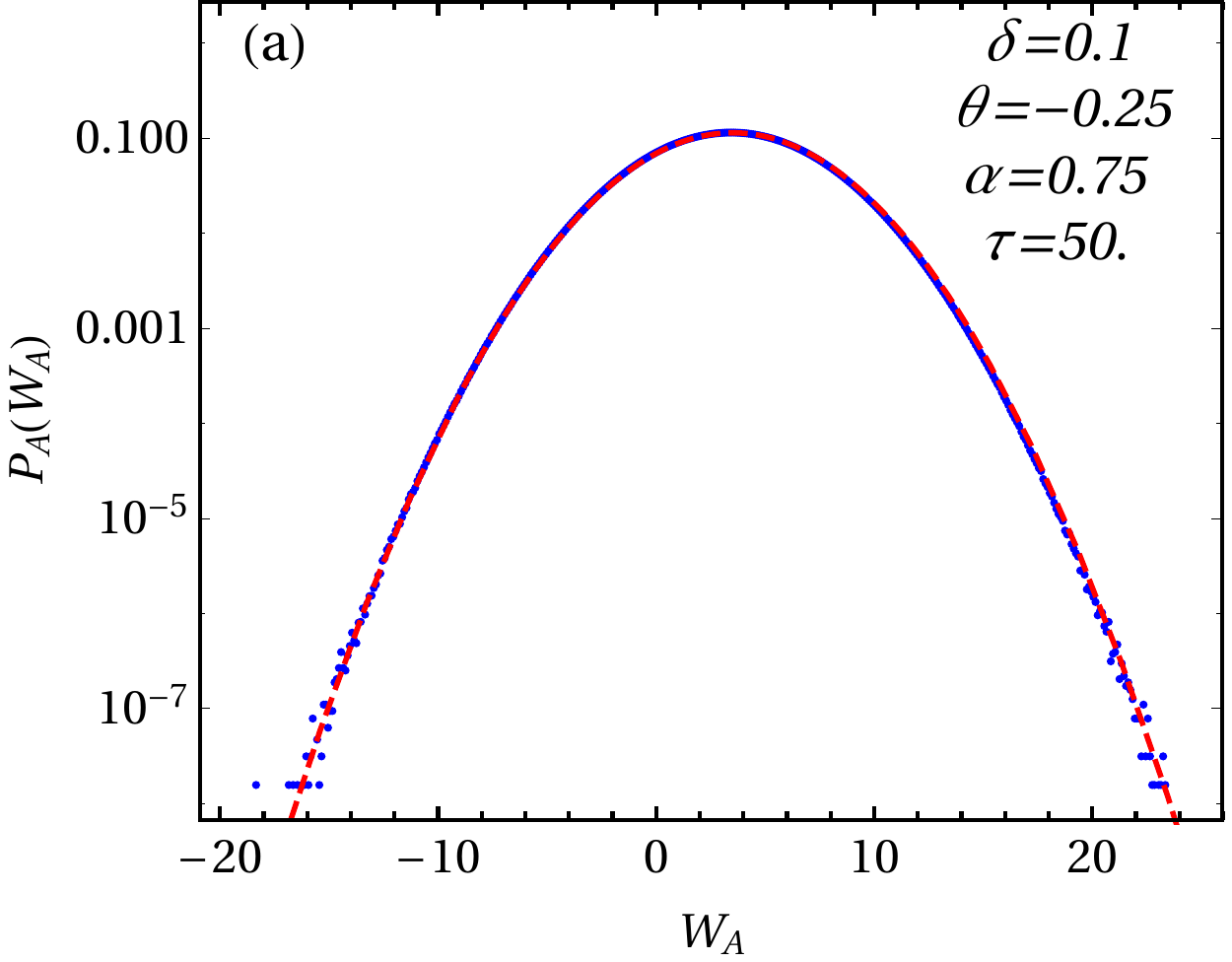}
    \includegraphics[width=6.7cm]{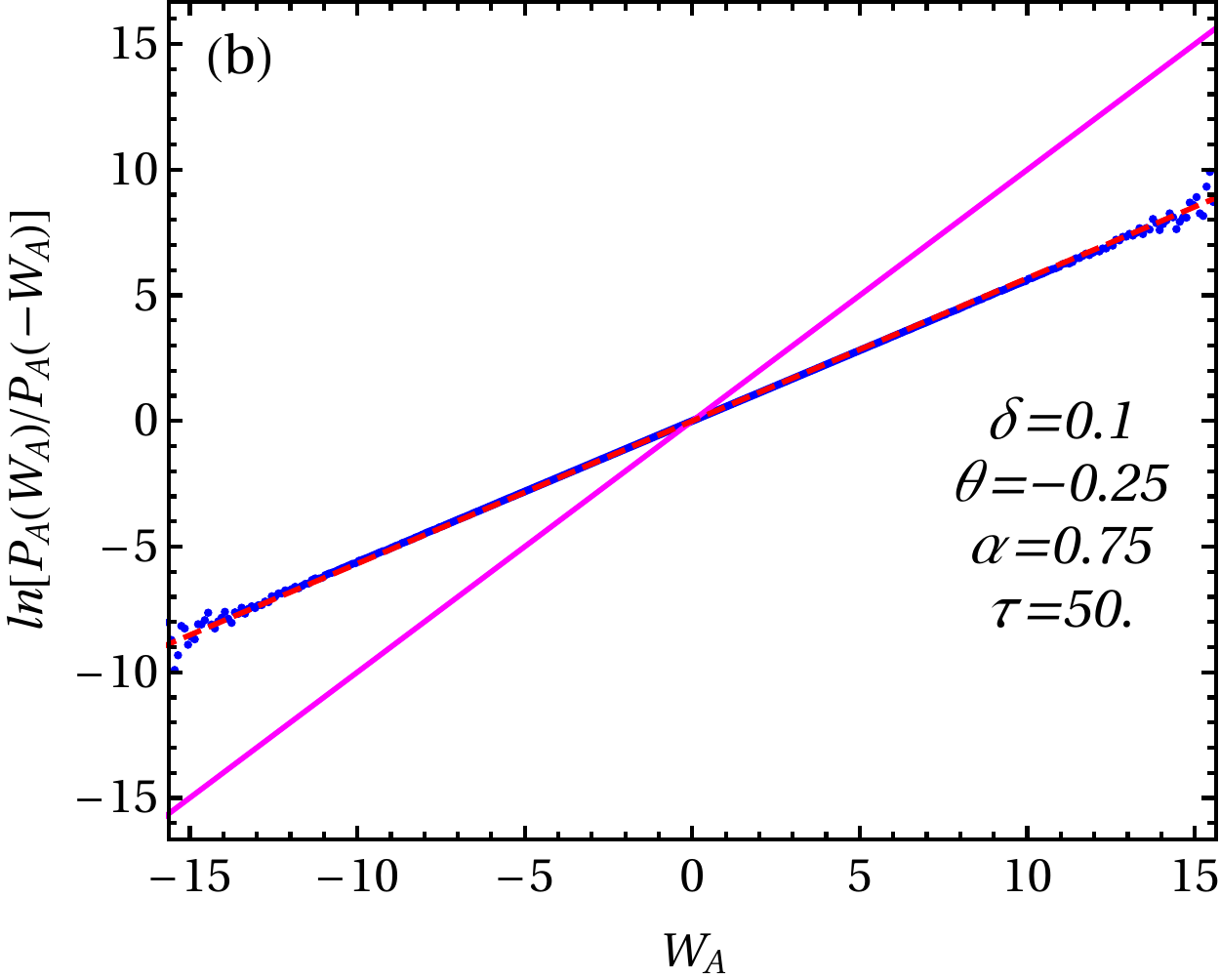}\\
\includegraphics[width=7cm]{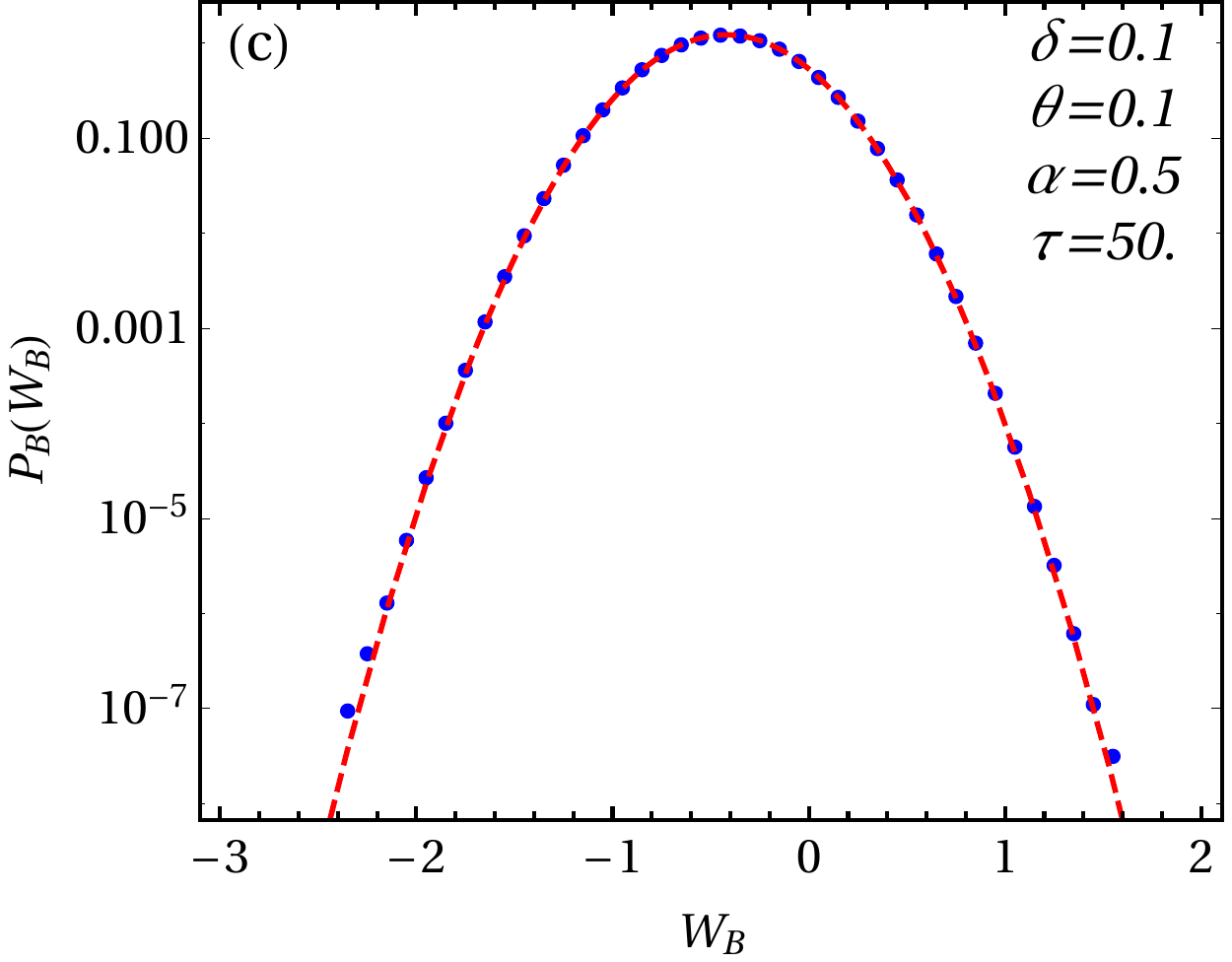}
    \includegraphics[width=6.7cm]{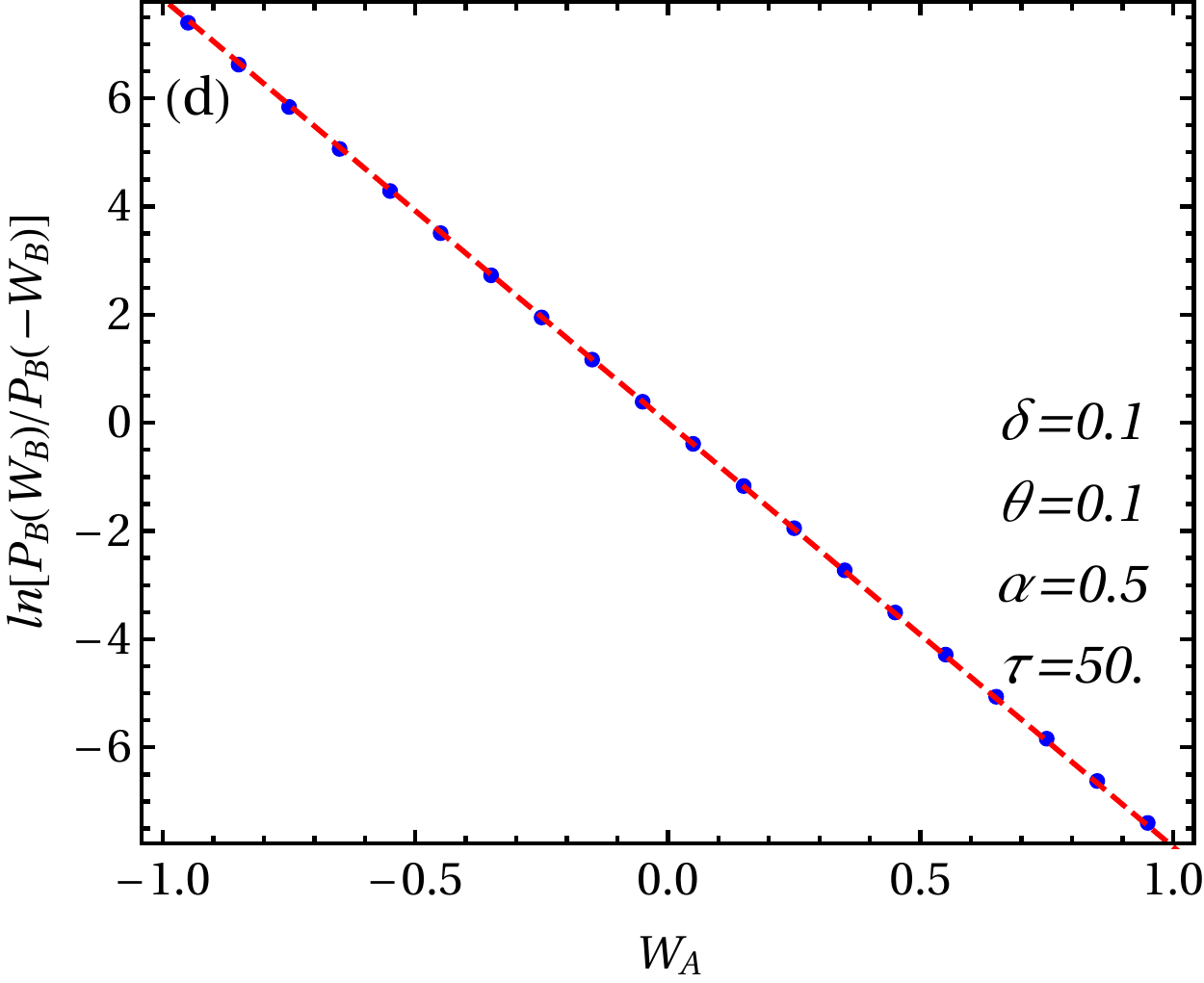}\\
    \end{tabular}
 \bigskip
   \caption{\label{small-delta-large-time} The probability density function for work done on particle A and B are shown in (a) and (c), respectively. The function $\ln[P_i(W_i)/P_i(-W_i)]$ are plotted against $W_i$ in (b) and (d). In all figures, the red dashed lines are the analytical result obtained from \eref{joint} with cumulants given in \erefss{mA-wc-lt}--\erefs{CBB-wc-lt} whereas the blue dots are obtained from numerical simulations. These results are shown for observation time $\tau\gg\lambda_2^{-1}$ and for fixed $\tau_\gamma=1$. Figures (b) and (d) indicate that probability density function for both work done ($W_A$ and $W_B$) do not satisfy TFT even in the weak coupling limit $(\delta\ll1)$ and for large observation time ($\tau\gg\lambda_2^{-1}\sim\delta^{-1}$). }
\end{figure*}

In \frefss{small-delta-small-time} (a) and \frefs{small-delta-small-time}(c), the comparison between the analytical probability density function for work done on particle A and B given in \eref{joint} in which the cumulants $(\mu_A,\mu_B,C_{AA},C_{BB})\to(\tilde{\mu}_A,\tilde{\mu}_B,\tilde{C}_{AA},\tilde{C}_{BB})$ are given in \erefss{mA-wc-st}--\erefs{cBB-wc-st},  with the numerical simulation  is shown for coupling strength $\delta=0.001$ and observation time $\tau=10\ll\delta^{-1}$. TFT for work done on particle A and B is shown in \frefss{small-delta-small-time}(b) and \frefs{small-delta-small-time}(d) for respective parameters. In all plots, blue points are the obtained from numerical simulations and the red dashed lines are the analytical results. In \frefss{small-delta-small-time}(b) and \frefs{small-delta-small-time}(d), red dashed lines have slope unity. Therefore, in the weak coupling limit $(\delta\ll 1)$ and observation time $u_\tau\ll\delta^{-1}$, TFT is satisfied. 

From \aref{Green-mat}, we see that $\lambda_1>\lambda_2>0$. Thus, in the large time limit ($u_\tau\gg\lambda_2^{-1}$), the cumulants for both work done ($W_A$ and $W_B$) simplify to 
\begin{align}
\bar{\mu}_A=&\dfrac{\alpha^2 \tau_\gamma}{2\delta}[2(1+\delta)+\theta(\delta u_\tau-2\delta-1)]\label{mA-wc-lt},\\
\bar{\mu}_B=&\dfrac{\alpha^2\theta \tau_\gamma}{4\delta}[2-\theta-2\delta(1-\theta)(u_\tau-2)]\label{mB-wc-lt},\\
\bar{C}_{AA}=&\dfrac{2(1+\delta)\alpha^2\tau_\gamma}{\delta}\label{CAA-wc-lt},\\
\bar{C}_{BB}=&\dfrac{\alpha^2\theta^2\tau_\gamma}{2\delta}[2\delta(u_\tau-2)-1]\label{CBB-wc-lt},
\end{align} 
where $u_\tau=\tau/\tau_\gamma.$
Notice that the above given results are true for all $\delta$ and for large observation time $u_\tau\gg\lambda_2^{-1}$.

In \frefss{small-delta-large-time}(a) and \frefs{small-delta-large-time}(c), we have shown a comparison of analytical probability density function for work done on particle A and particle B, respectively, given in \eref{joint} in which cumulants $(\mu_A,\mu_B,C_{AA},C_{BB})\to(\bar{\mu}_A,\bar{\mu}_B,\bar{C}_{AA},\bar{C}_{BB})$ are given as \erefss{mA-wc-lt}--\erefs{CBB-wc-lt}, with the numerical simulation for coupling strength $\delta=0.1$ and time $\tau=50 \gg\delta^{-1}$. The variations of function $\ln[P_i(W_i)/P_i(-W_i)]$ are shown against $W_i$ in \frefss{small-delta-large-time}(b) and \frefs{small-delta-large-time}(d) where magenta solid line in \fref{small-delta-large-time}(b) has slope unity. Thus, in the weak coupling limit $(\delta\to0)$, $\lambda_2\sim \delta$, the violation of TFT can be seen for both work done ($W_A$ and $W_B$) for large observation time limit $u_\tau\gg\delta^{-1}$.

In above two cases, we have studied the TFT for $W_A$ and $W_B$ in the weak coupling limit ($\delta\to 0$) and showed that TFT would hold in the limit $u_\tau\ll \delta^{-1}$ whereas violation can be seen in the large observation time $u_\tau\gg\delta^{-1}$. This is because in the small time limit, the effect from the other particle will not affect the TFT of the work done on the observed particle. However, when the time of observation is large $(u_\tau\gg\delta^{-1})$, the effect from the other particle may appear  as the term coupling ($\delta$) times relative separation $(x_A-x_B)$ becomes relevant [see \eref{dyn-1} and \eref{dyn-2}] which leads to the violation of TFT of the work done of the observed particles even in the weak coupling limit. Similar results one can see in Refs. \cite{Deepak,HT}. 

%%%%%%%%%%%%%%%%%%%%%%%%%%%%%%%%%%%%%%%%%%%%%%%%%%%%%%%%%%%%%%%%%%%%%%%%%%%%%%%%%%%
%%%%%%%%%%%%%%%%%%%%%%%%%%%%%%%%%%%%%%%%%%%%%%%%%%%%%%%%%%%%%%%%%%%%%%%%%%%%%%%%%%%
\section{Transient fluctuation theorem for total work done on coupled system}
\label{tft}
In the above section, we have analyzed TFT for work done on particle A and B. In this section, we study the TFT for the total work done on both particles.

Total work done on particle A and B is given by
\begin{equation}
W=W_A+W_B,
\end{equation}
Since $W_A$ and $W_B$ have Gaussian distribution, therefore, $W$ also has Gaussian distribution with mean $\mu=\overline{\langle W \rangle}$ and variance $C=\overline{\langle [W-\overline{\langle W \rangle}]^2 \rangle}$ given as
\begin{align}
\mu&=\mu_A+\mu_B,\\
C&=C_{AA}+C_{BB},
\end{align}
where $C=2 \mu$ and 
\begin{align}
\mu&=\dfrac{\alpha^2\tau_\gamma}{4\delta}\bigg[4(1+\delta)-[1+2\delta(2-u_\tau)]\theta^2-e^{-(1+2 \delta)u_\tau/2}\bigg\{\big[4-\theta+4\delta(1-\theta^2)\big]\nonumber\\&\times\cosh\big[\sqrt{1+4\delta^2}u_\tau/2\big]+\dfrac{\big[4-\theta^2+2\delta\{2-\theta^2+4\delta(1-\theta^2)\}\big]}{\sqrt{1+4\delta^2}}\sinh\big[\sqrt{1+4\delta^2}u_\tau/2\big]\bigg\}\bigg].
\end{align}

Therefore, the probability density function for total work done $W$ is given by
\begin{equation}
P(W)=\dfrac{1}{\sqrt{2 \pi C}} \exp\bigg[-\dfrac{(W-\mu)^2}{2 C}\bigg].
\label{joint-W}
\end{equation}
Using the above relation, one can see that $P(W)/P(-W)=e^W$, \emph{i.e.,} TFT for total work done on both particles is satisfied. Therefore, it is clear that when degrees of freedom (DOFs) having same time of relaxation are coupled and driven out of equilibrium, total work done on all DOFs obeys the TFT for all parameters.
\begin{figure*}
 \begin{tabular}{cc}
    \includegraphics[width=7cm]{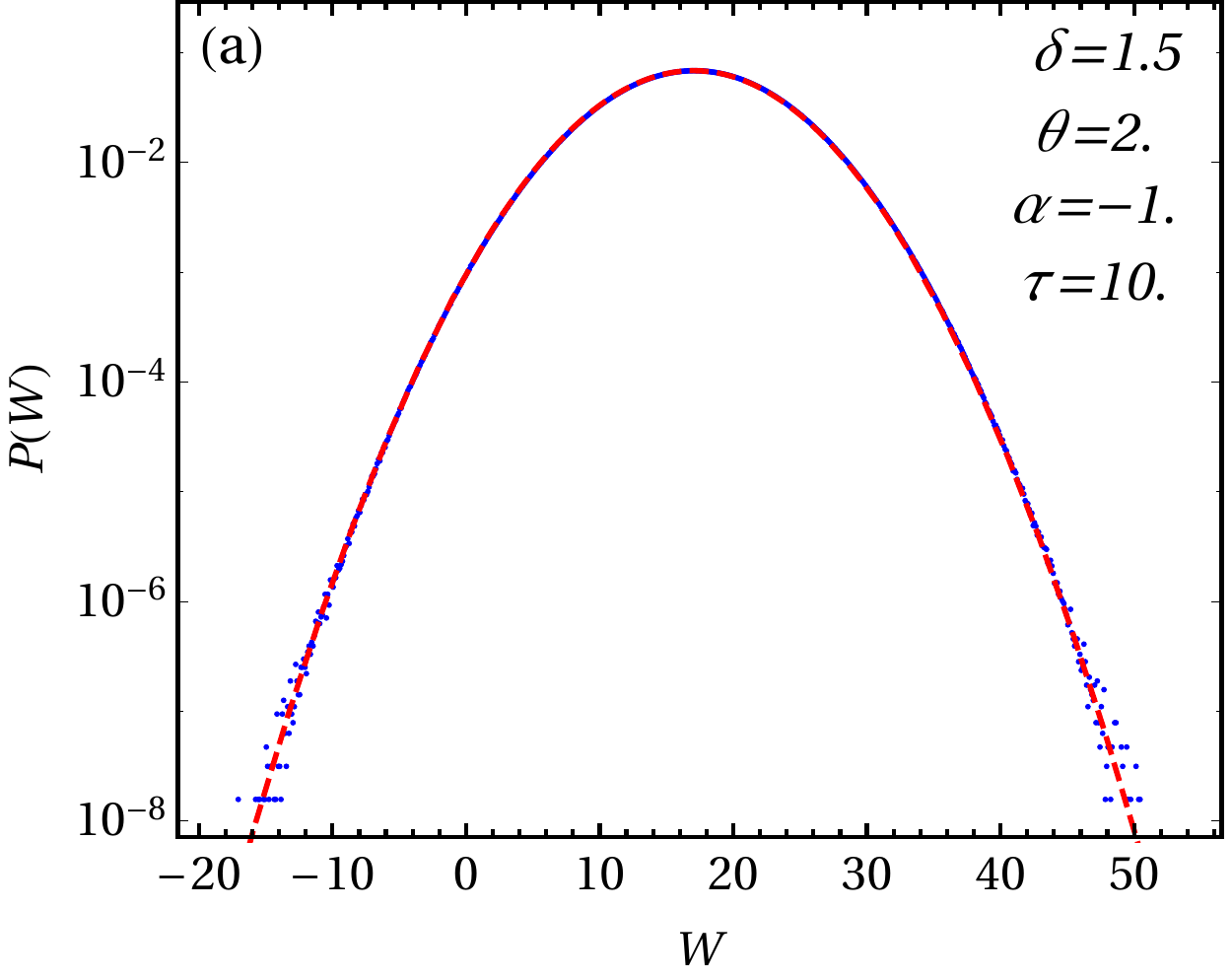}
    \includegraphics[width=7cm]{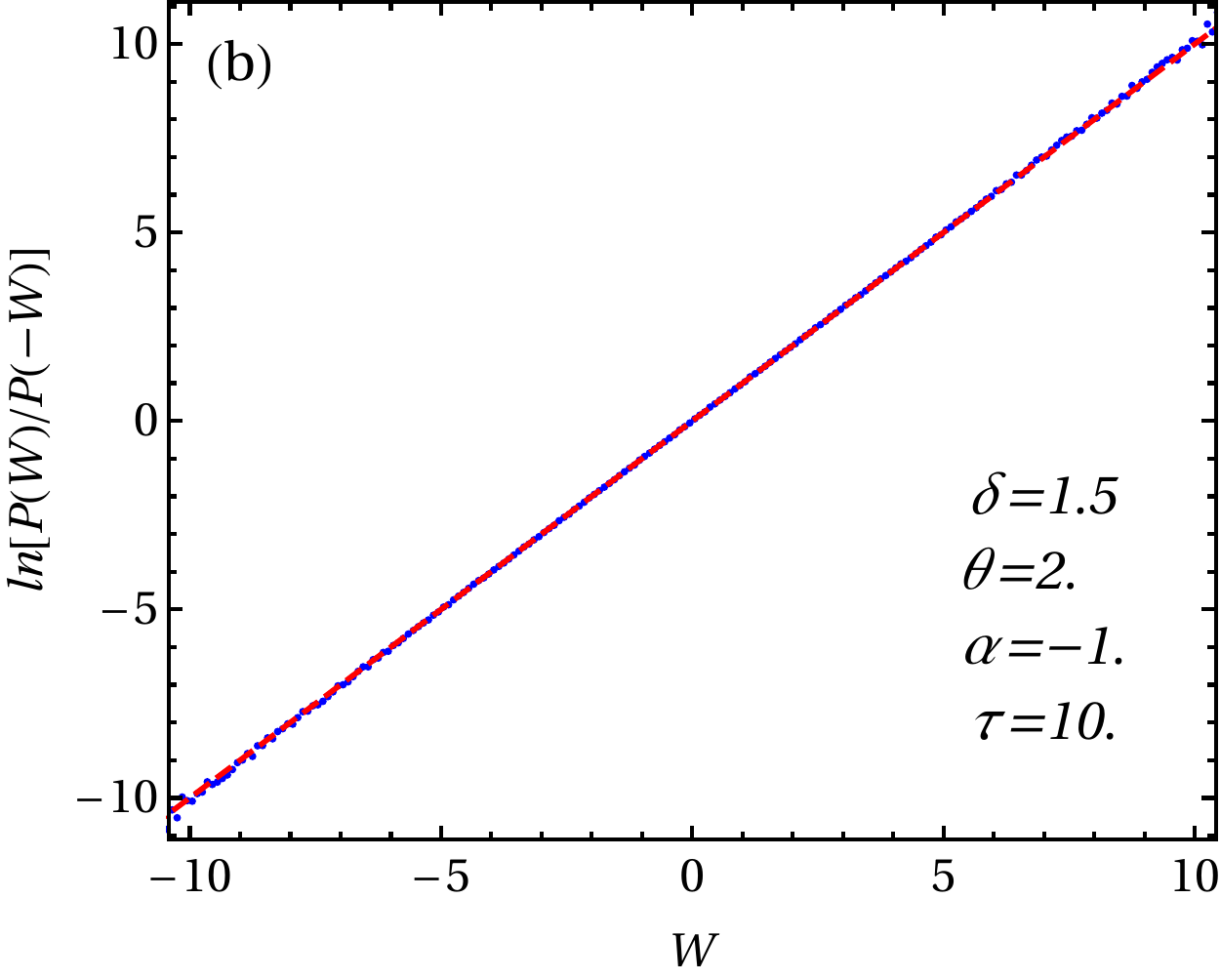}
    \end{tabular}
 \bigskip
   \caption{\label{pdf-Wtot} (a) The distribution function given in \eref{joint-W} is compared with the numerical simulation. (b) The function $\ln[P(W)/P(-W)]$ is plotted against the $W=W_A+W_B$. In both figures, the parameters are  $\delta=1.5$, $\theta=2.0$, $\alpha=-1.0$ and $\tau=10$, and the red dashed line is the analytical result whereas the blue dots are obtained from the numerical simulation. The red dashed line in (b) has unit slope. These results are shown for $\tau_\gamma=1$.}
\end{figure*}

\ffref{pdf-Wtot}(a) and \frefs{pdf-Wtot}(b) show the comparison of the analytical results for the probability density function for $W$ given in \eref{joint-W} and the function $\ln[P(W)/P(-W)]$ with the numerical simulation, and they have nice agreement. \ffref{pdf-Wtot}(b) indicates that the total work done $W$ obeys the transient fluctuation theorem.

%%%%%%%%%%%%%%%%%%%%%%%%%%%%%%%%%%%%%%%%%%%%%%%%%%%%%%%%%%%%%%%%%%%%%%%%%%%%%%%%%%%%%%%%%%%%%%%%%%%%%%%%%%%%%%%%%%%%%%%%%%%%%%
%%%%%%%%%%%%%%%%%%%%%%%%%%%%%%%%%%%%%%%%%%%%%%%%%%%%%%%%%%%%%%%%%%%%%%%%%%%%%%%%%%%%%%%%%%%%%%%%%%%%%%%%%%%%%%%%%%%%%%%%%%%%%%%
\section{Probability density function for stochastic efficiency $p_\tau(\eta)$}
\label{p-eta-sec}
The main objective of the paper is to understand the statistics of the stochastic efficiency. Therefore, substituting $P_A(W_A)$ and $P_B(W_B)$ given in \eref{joint}, in the integral \eref{p-eta-int}, the probability density function for stochastic efficiency $p_\tau(\eta)$ can be obtained as  
\begin{equation}
p_\tau(\eta)=\dfrac{e^{I(\eta,\tau)}}{\sqrt{(2 \pi)^2 C_{AA}C_{BB}}}\dfrac{e^{-K_1K_2^2}+K_2\sqrt{\pi K_1}\ \mathrm{erf}(K_2\sqrt{K_1})}{K_1},
\label{p-eta}
\end{equation}
where 
\begin{align}
I(\eta,\tau)&=-\dfrac{1}{2}\dfrac{(\eta \mu_B +\mu_A)^2}{\eta^2 C_{BB}+C_{AA}},\label{LDF}\\
K_1&=\dfrac{\eta^2C_{BB}+C_{AA}}{2 C_{AA}C_{BB}},\label{K1}\\
K_2&=\dfrac{C_{AA}\mu_B-\eta C_{BB}\mu_A}{C_{AA}+\eta^2C_{BB}}\label{K2}.
\end{align}
In \eref{p-eta}, $\mathrm{erf}(u)$ is the error function given by
\begin{equation}
\mathrm{erf}(u)=\dfrac{2}{\sqrt{\pi}}\int_0^u dx\  e^{-x^2}.
\end{equation}

\begin{figure*}
 \begin{tabular}{ccc}
    \includegraphics[width=5.0cm]{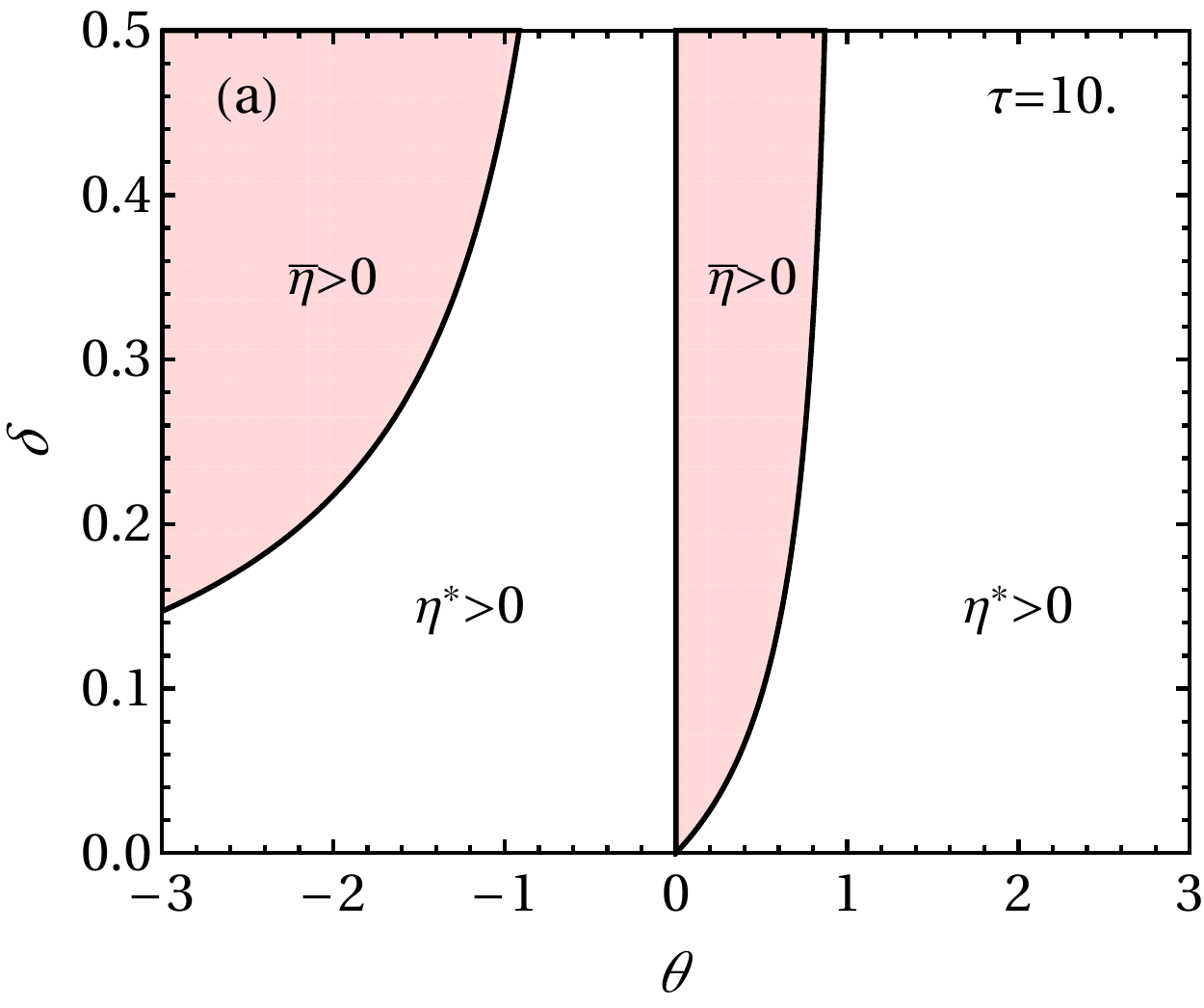}
    \includegraphics[width=5.0cm]{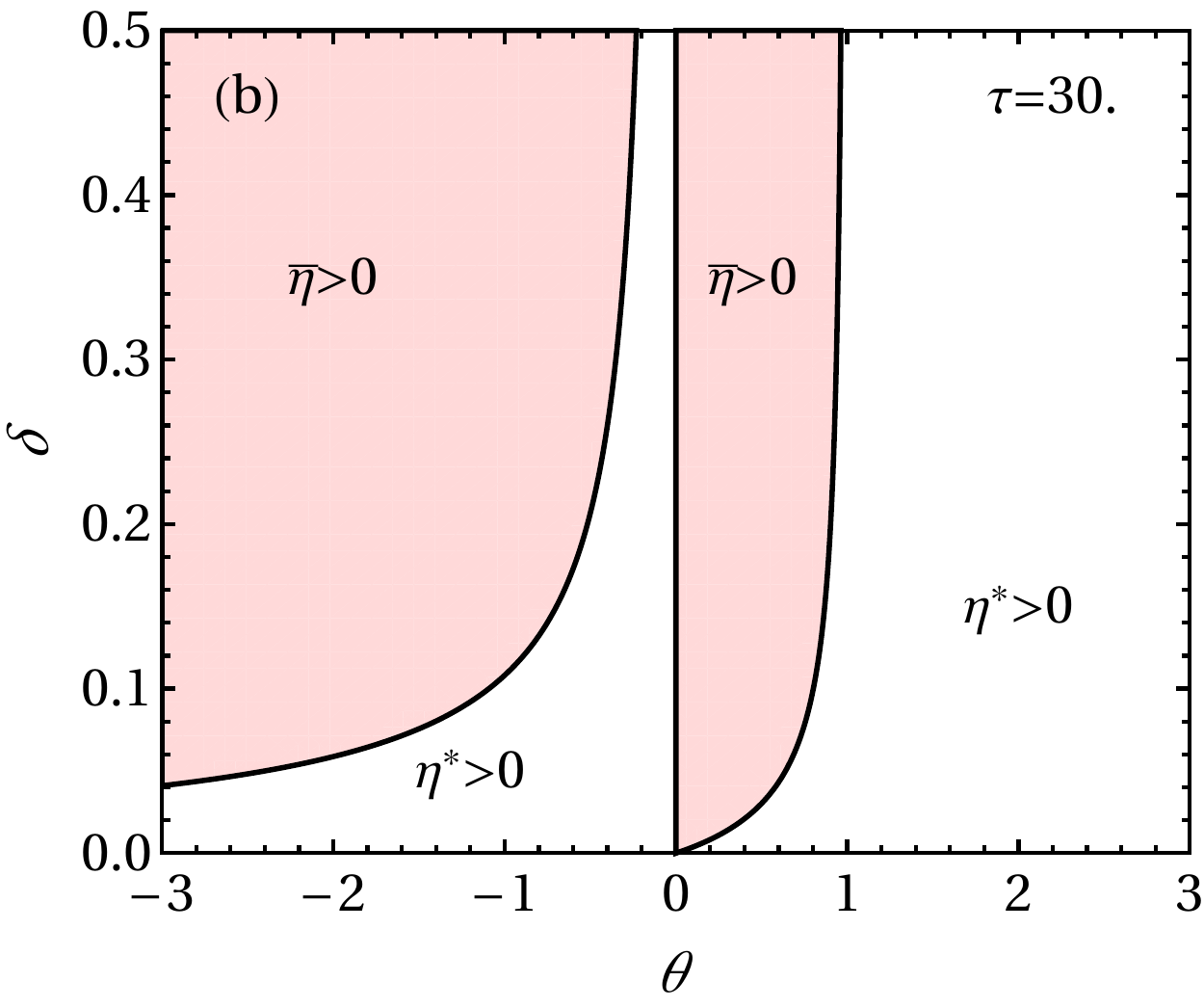}
    \includegraphics[width=5.0cm]{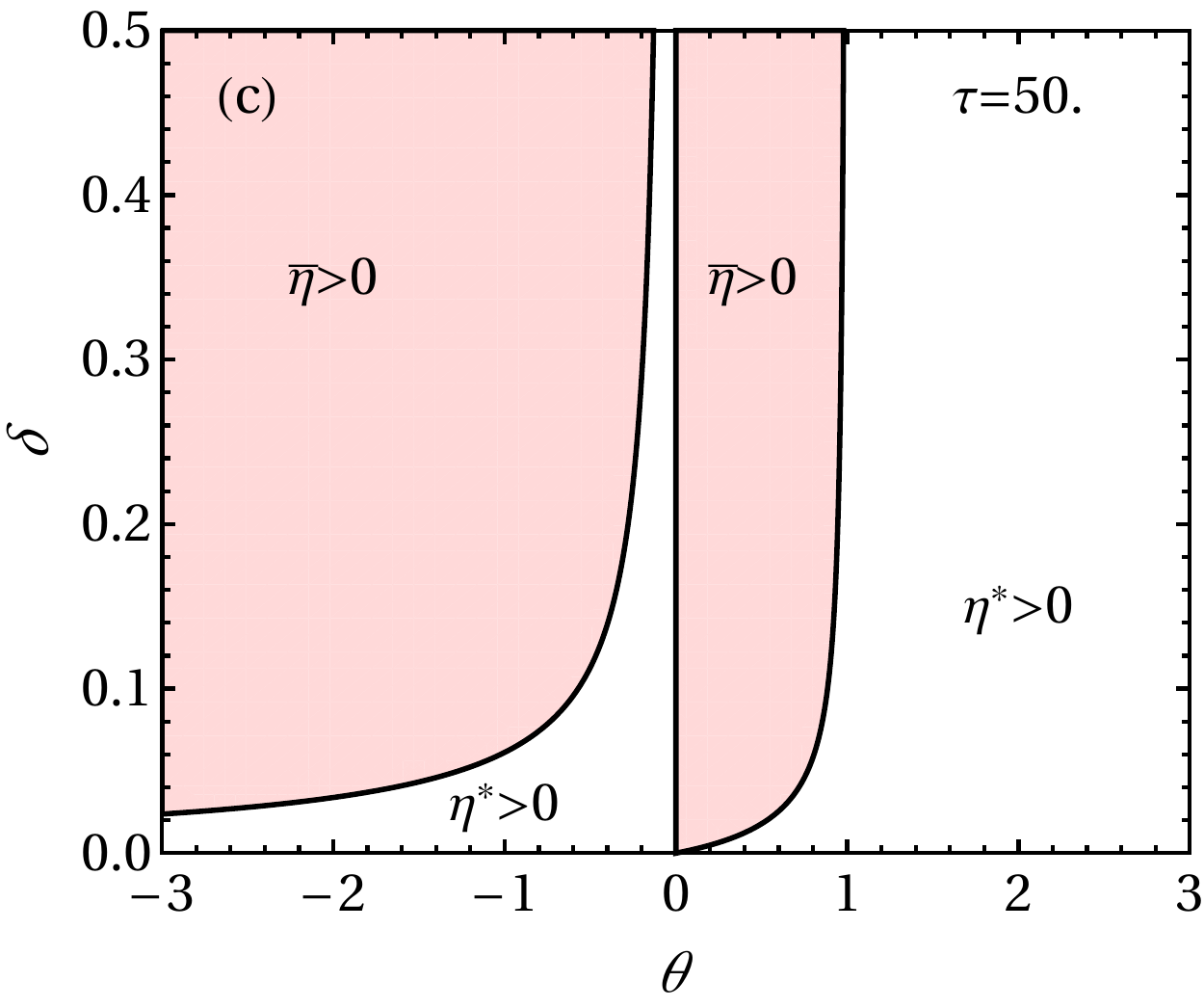}
    \end{tabular}
 \bigskip
 \caption{\label{phase-dig-time} Phase diagrams in $(\theta,\delta)$ plane are shown using $\bar{\eta}$ and $\eta^*$ given in \eref{extreme}, for given time $\tau$. The shaded regions correspond to the areas where the efficiencies $\bar{\eta}$ and $\eta^*$ remain positive and negative, respectively. The above phase diagrams are shown for given $\tau_\gamma=1$.}
\end{figure*}

It can be seen that the function $I(\eta,\tau)$ has two extrema, \emph{i.e.,} at $\bar{\eta}$ and $\eta^*$ where 
\begin{equation}
\bar{\eta}=-\dfrac{\mu_A}{\mu_B}, \quad \quad\quad   \eta^*=\dfrac{\mu_B C_{AA}}{\mu_A C_{BB}}.
\label{extreme}
\end{equation}
Moreover, $I(\eta,\tau)|_{\eta=\bar{\eta}}=0$.
The function $I(\eta,\tau)$ has a maximum and a minimum at $\eta=\bar\eta$ and $\eta=\eta^*$, respectively. 
Note that both $\bar{\eta}$ and $\eta^*$ do not depend upon the parameter $\alpha$ (relative strength acting on particle A, \emph{i.e.,} $\alpha=F/\sqrt{\gamma T}$).
The efficiency $\bar{\eta}$ can have any sign depending upon the parameters $\tau$, $\delta$, and $\theta$. To understand the nature of $\bar{\eta}$, we have plotted phase diagram in $(\theta,\delta)$ plane as shown in \fref{phase-dig-time} for fixed time $\tau$. In \fref{phase-dig-time}, the light red shaded regions correspond to the areas where the efficiency $\bar{\eta}$ is positive. Note that $C_{AA}$ and $C_{BB}$ are positive [see \eref{extreme}]. Therefore, $\bar{\eta}$ and $\eta^*$ have opposite sign. Hence, the unshaded regions in \fref{phase-dig-time} represent the areas where $\eta^*$ is positive otherwise negative. 
\begin{figure*}
 \begin{tabular}{cc}
    \includegraphics[width=7cm]{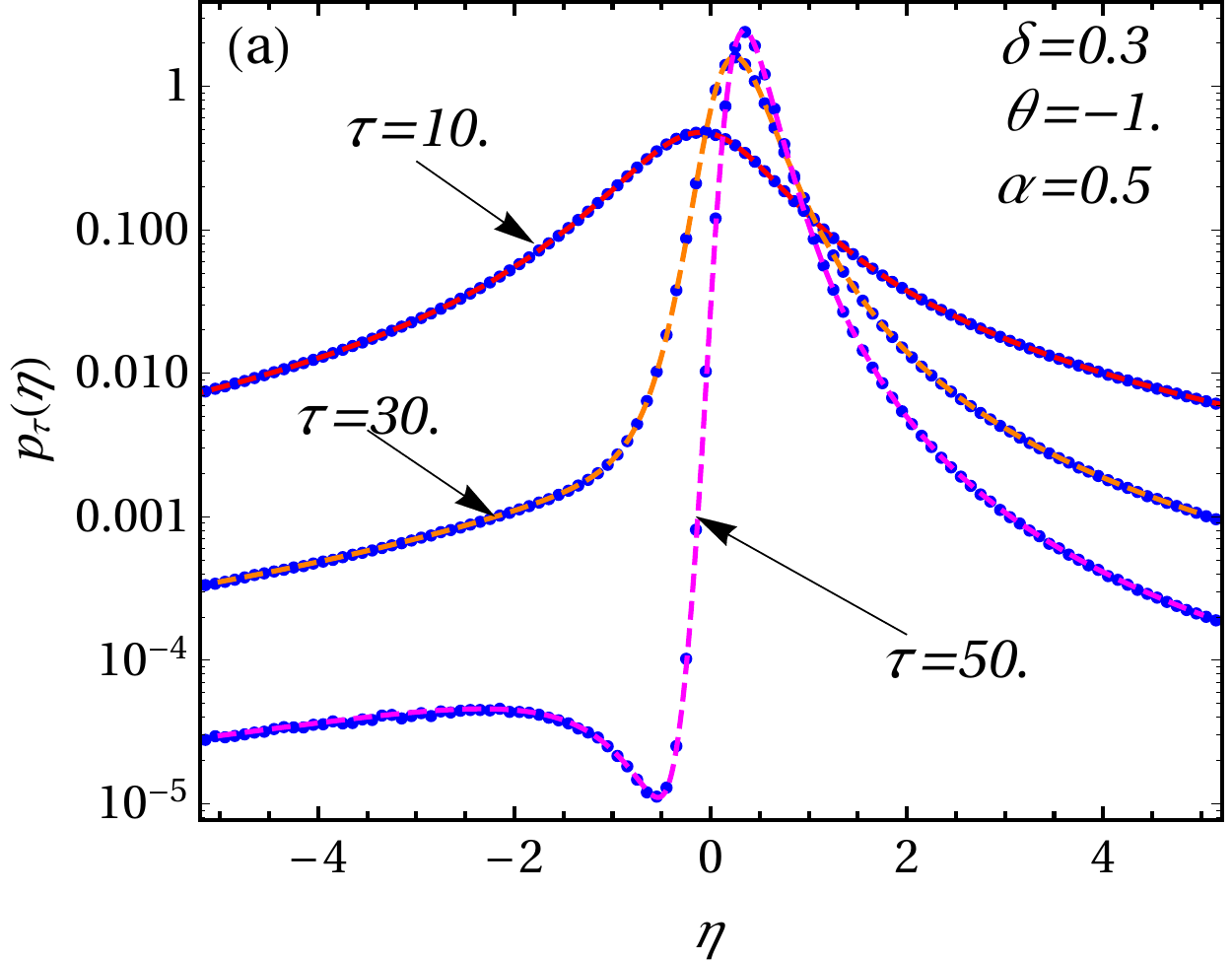}
    \includegraphics[width=7cm]{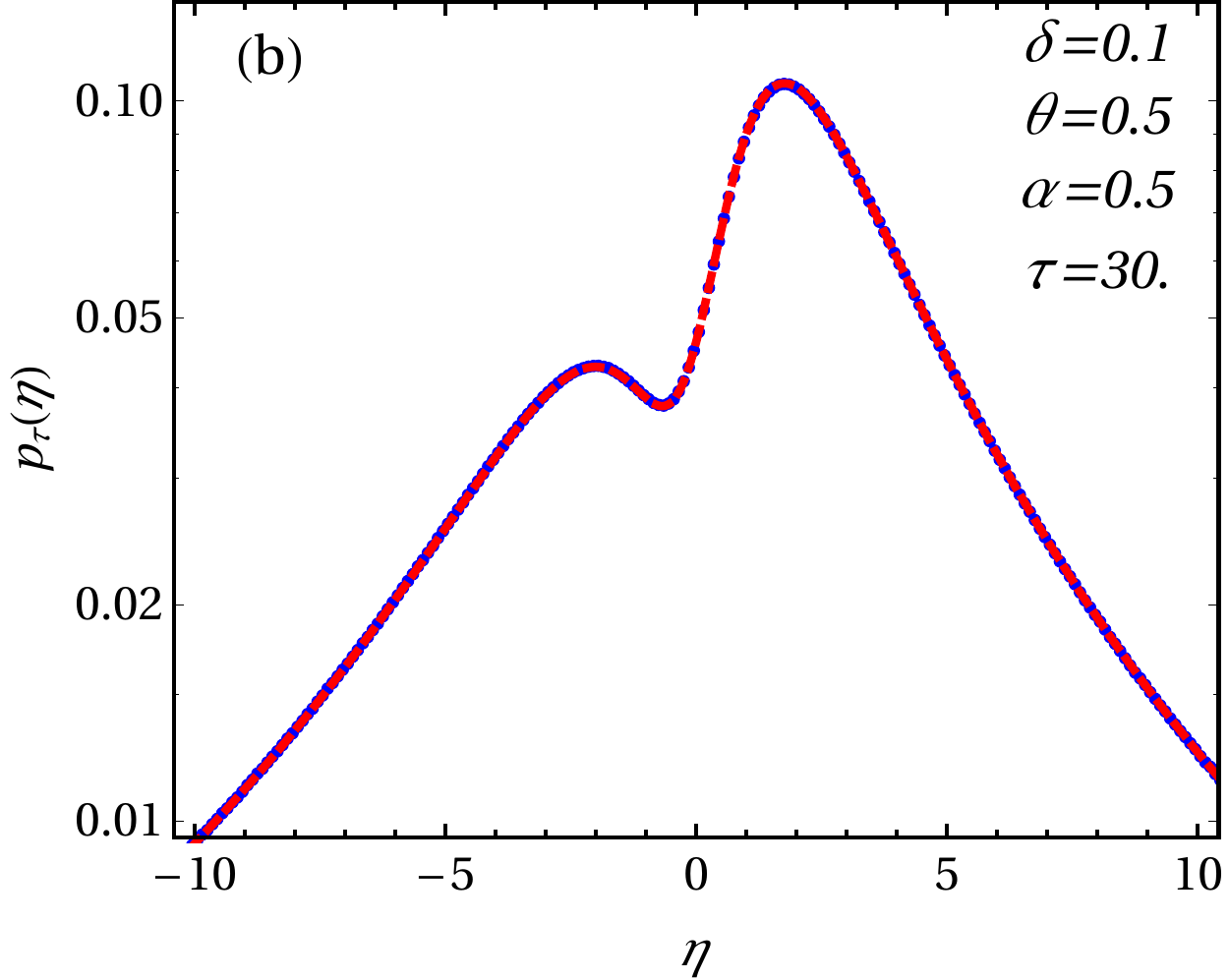}
    \end{tabular}
 \bigskip
   \caption{\label{pdf-prob}The probability density function for stochastic efficiency $p_\tau(\eta)$ is plotted against the stochastic efficiency $\eta$: (a) for $\theta=-1.0$, $\alpha=0.5$, coupling parameter $\delta=0.3$ at three different time $\tau=10$, $\tau=30$ and $\tau=50$ where red ($\tau=10$), orange ($\tau=30$) and magenta ($\tau=50$) dashed lines represent the analytical results given by \eref{p-eta}, and (b) for $\theta=0.5$, $\alpha=0.5$, coupling parameter $\delta=0.1$ at time $\tau=30$ where red dashed lines represent the analytical result given in \eref{p-eta}. In both figures, blue dots are obtained from the numerical simulations at the respective times $\tau$. These results are shown for fixed $\tau_\gamma=1$. }
\end{figure*}

\ffref{pdf-prob} shows the comparison of analytical results of the probability density function $p_\tau(\eta)$ given in \eref{p-eta} with the numerical simulation results for both signs of $\theta$.  While in \fref{pdf-prob}(a) the comparison is shown for parameters $\theta=-1.0$,  $\alpha=0.5$, and coupling parameter $\delta=0.3$ at $\tau=10$, $\tau=30$ and $\tau=50$, the parameters $\theta=0.5$, $\alpha=0.5$, and coupling parameter $\delta=0.1$ at $\tau=30$ are taken in \fref{pdf-prob}(b). In \fref{pdf-prob}(a), the red ($\tau=10$), orange ($\tau$=30) and magenta ($\tau$=50) dashed lines are the analytical result given by \eref{p-eta}  whereas the blue dots are obtained from the numerical simulations at corresponding times $\tau$. Similarly, the red dashed lines correspond to the analytical result given by \eref{p-eta} and blue dots are obtained from numerical simulation in \fref{pdf-prob}(b). Both of figures show that there is a nice agreement between theory and numerical simulation. From \fref{pdf-prob}(a), it is clear that the peak of the density function $p_\tau(\eta)$ shifts from negative to positive side as time $\tau$ is increased from $\tau=10$ to $\tau=30$ as shown in \frefss{phase-dig-time}(a)---\frefs{phase-dig-time}(c).

\begin{figure*}
 \begin{tabular}{cc}
    \includegraphics[width=7cm]{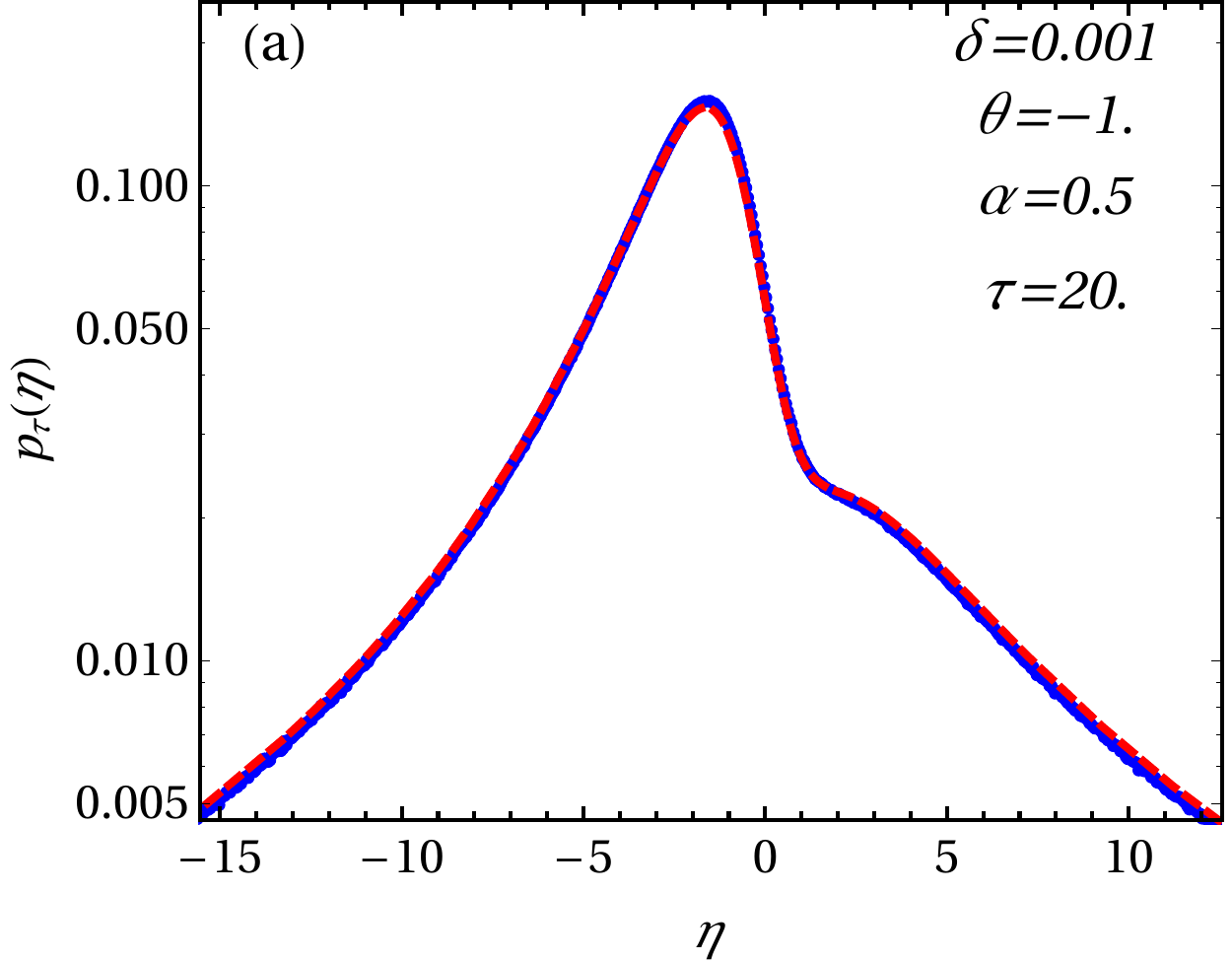}
    \includegraphics[width=7cm]{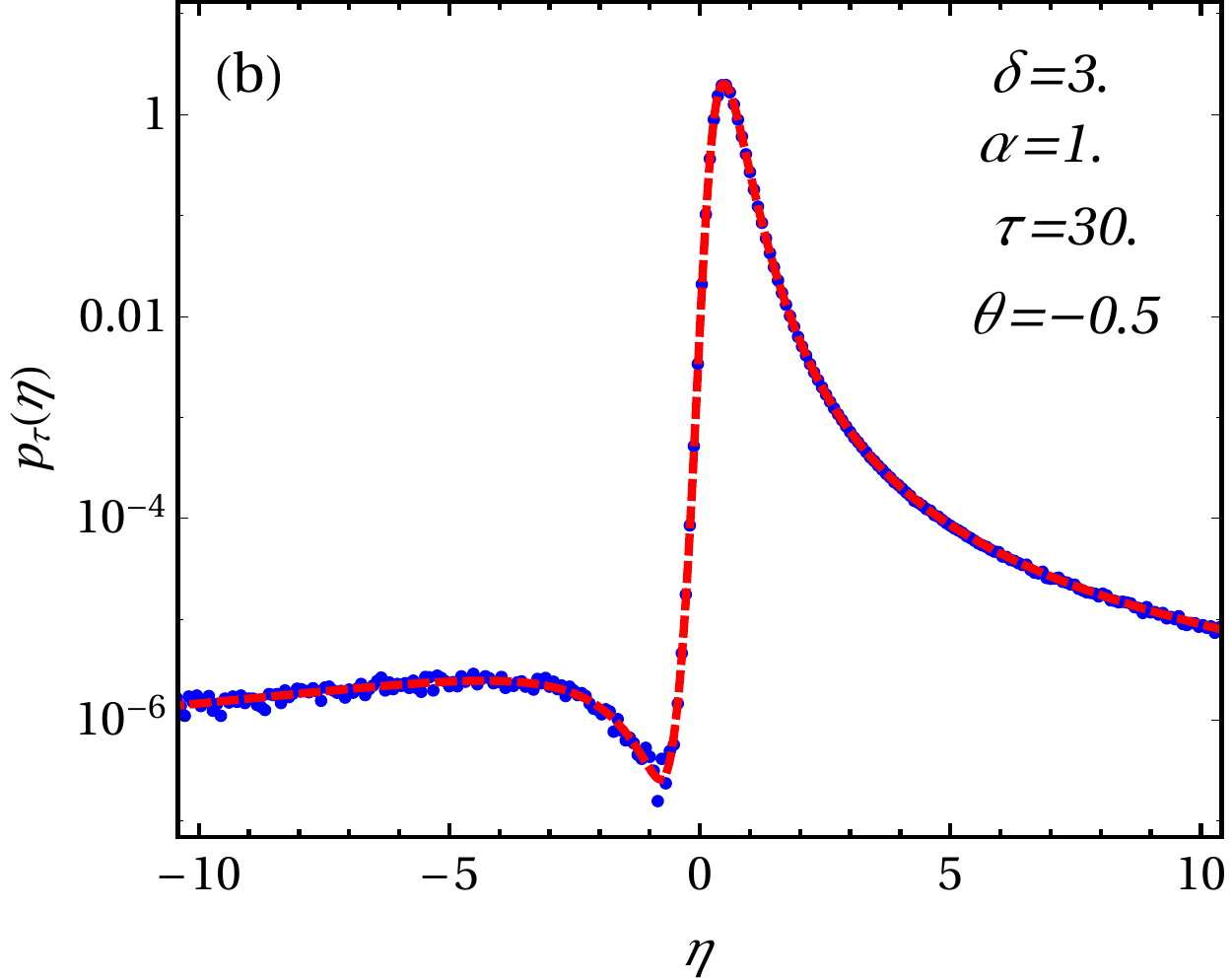}
    \end{tabular}
 \bigskip
   \caption{\label{large-small-eta-prob}The probability density functions for the stochastic efficiency $p_\tau(\eta)$ are plotted against the stochastic efficiency $\eta$ for weak coupling $(\delta\ll 1)$ and for small observation time $(\tau \ll\delta^{-1})$ [see (a)], and for large observation time $\tau\gg\lambda_2^{-1}$ [see (b)]. In both figures, blue dots are obtained from numerical simulations and red dashed lines are the analytical probability density function. We set $\tau_\gamma=1$ for both above plots. }
\end{figure*}

In the above, we have given efficiencies ($\bar{\eta}$ and $\eta^*$) where $I(\eta,\tau)$ has extrema. It can be seen that in the weak coupling limit $(\delta\to 0)$ and small time limit ($u_\tau\ll\delta^{-1}$), these efficiencies reduce to 
\begin{align}
\bar{\eta}&=-\dfrac{4}{\theta^2}\dfrac{u_\tau}{e^{-u_\tau}+u_\tau-1},\\
\eta^*&=1.
\end{align} 
Clearly, here the most probable efficiency $\bar{\eta}$ is negative. This is because in the small coupling limit ($\delta\to 0$) and small observation time ($u_\tau \ll \delta^{-1}$), both of the particles behave independently [see \fref{small-delta-small-time}], and the machine will not do work against the load force $F$ irrespective the value of $\theta$ (in most probable sense). In \fref{large-small-eta-prob}(a), we have plotted the probability density function for the stochastic efficiency $p_\tau(\eta)$ in the weak coupling limit and small time of observation where the red dashed lines represents the analytical results given by \eref{p-eta} in which $(\mu_A,\mu_B,C_{AA},C_{BB})\to(\tilde{\mu}_A,\tilde{\mu}_B,\tilde{C}_{AA},\tilde{C}_{BB})$ as given in \erefss{mA-wc-st}--\erefs{cBB-wc-st} and blue dots correspond to the numerical  simulation results. 
Similarly, one can obtain $\bar{\eta}$ and $\eta^*$ in the large time limit $(u_\tau\gg\lambda_2^{-1})$ from \erefss{mA-wc-lt}--\erefs{CBB-wc-lt}
\begin{align}
\bar{\eta}&=-\frac{2 [\delta  \{\theta  (u_\tau -2)+2\}-\theta+2]}{\theta  [2 \delta  (\theta -1) (u_\tau -2)-\theta +2]},\label{eta-bar-lt}\\
\eta^*&=\frac{2 (\delta +1) [2 \delta  (\theta -1) (u_\tau -2)-\theta +2]}{\theta  [2 \delta  (u_\tau -2)-1] [\delta  \{\theta  (u_\tau -2)+2\}-\theta +2]}\label{eta-s-lt}.
\end{align}

We have also compared the large time ($u_\tau\gg\lambda_2^{-1}$) analytical results for the probability density function $p_\tau(\eta)$ given in \eref{p-eta} in which $(\mu_A,\mu_B,C_{AA},C_{BB})\to(\bar{\mu}_A,\bar{\mu}_B,\bar{C}_{AA},\bar{C}_{BB})$ as given in \erefss{mA-wc-lt}--\erefs{CBB-wc-lt},  with the numerical simulations in \fref{large-small-eta-prob}(b). These comparisons indicate that there is a nice agreement between theory and numerical simulations.

In large time limit ($u_\tau\gg\lambda_2^{-1}$), when $\theta=1$, we see that 
\begin{align}
\bar{\eta}&=-2(1+\delta u_\tau),\\
\eta^*&=\dfrac{2(1+\delta)}{(1+\delta u_\tau)[2\delta(u_\tau-2)-1]}.
\end{align}
However, \eref{eta-bar-lt} and \eref{eta-s-lt} reduce to $\bar{\eta}\to 1/(1-\theta)$ and $\eta^*\to 0$ for $\theta\neq1$ as $u_\tau\to \infty$.
Therefore, the machine will not perform work against the load force $F$ (in most probable sense) when $\theta\geq 1$ in the large time limit as $\bar{\eta}<0$ .

Finally, we emphasize that for all cases shown above, probability density function for the stochastic efficiency $p_\tau(\eta)\to\eta^{-2}$ as $|\eta|\to\infty$. Similar behaviour of $p_\tau(\eta)$ has been observed earlier in different model systems \cite{Gingrich,Proesmans-1,Polettini,Proesmans-3,Deepak-eff}. 

\section{Summary}
\label{summ-sec}
We considered two Brownian particle (say particle A and B) interacting harmonically with a spring of stiffness $k$. Particle B is confined in a harmonic trap of stiffness $k_0$. For simplicity, we defined a dimensionless coupling parameter $\delta$ as the ratio of spring constant and trap strength: $\delta=k/k_0$.  For $t\leq0$, the harmonic trap was kept  stationary. Thus, the system obeyed the equilibrium Boltzmann's distribution at $t=0$. When $t>0$, a constant load $F=\alpha \sqrt{\gamma T}$ is attached to particle A and the minimum of the harmonic confinement is dragged with a constant velocity $v=F\theta/(2\gamma)$. The joint distribution of both work done is computed. Transient fluctuation theorem (TFT) is studied for work done on particle A and B in the weak coupling limit $\delta\ll 1$ for both small ($u_\tau\ll\delta^{-1}$) and large observation time ($u_\tau\gg\delta^{-1}$). It is shown that the TFT would hold in the small time limit whereas the violation can be seen even in the weak coupling limit ($\delta\ll 1$) for large time.  Interestingly, the total work done on both particles satisfies TFT for all parameters. Further, we computed the stochastic efficiency which is defined as the work done against the load on particle A to the work done on the particle B by dragging the harmonic confinement. The exact distribution for the stochastic efficiency is evaluated for all time $\tau$ and the coupling parameter $\delta$. We have given the phase diagrams in $(\theta,\delta)$ plane for given time $\tau$ which shows the region where the efficiency at which the probability density is maximum attains the positive value. The analytical results are also supported by the numerical simulation and they have an excellent match.

As a final remark, this model system can be realized in an experiment \cite{Wang,Wang-2,Ribezzi-Crivellari}, and it would be interesting to compare the experimental results with the theoretical predictions.

\section*{Acknowledgement}
The author thanks Sanjib Sabhapandit for useful discussions.

%%%%%%%%%%%%%%%%%%%%%%%%%%%%%%%%%%%%%%%%%%%%%%%%%%%%%%%%%%%%%%%%%%%%%%%%%%%%
%%%%%%%%%%%%%%%%%%%%%%%%%%%%%%%%%%%%%%%%%%%%%%%%%%%%%%%%%%%%%%%%%%%%%%%
%% This part is added (a modification in the iopart.cls) to modify the
%% tableofcontents entry for the appendices
%%%%%%%%%%%%%%%%%%%%%%%%%%%%%%%%%%%%%%%%%%%%%%%%%%%%%%%%%%%%%%%%%%%%%%%
\makeatletter%
\def\@sect#1#2#3#4#5#6[#7]#8{\ifnum #2>\c@secnumdepth
  \let\@svsec\@empty\else
  \refstepcounter{#1}\edef\@svsec{\csname the#1\endcsname. }\fi
  \@tempskipa #5\relax
  \ifdim \@tempskipa>\z@
  \begingroup #6\relax
  \noindent{\hskip #3\relax\@svsec}{\interlinepenalty \@M #8\par}%
  \endgroup
  \csname #1mark\endcsname{#7}\addcontentsline
          {toc}{#1}{\ifnum #2>\c@secnumdepth \else
            \protect\numberline{}{\csname the#1\endcsname}\fi
            . #7}\else
          \def\@svsechd{#6\hskip #3\relax  %% \relax added 2 May 90
            \@svsec #8\csname #1mark\endcsname
                    {#7}\addcontentsline
                    {toc}{#1}{\ifnum #2>\c@secnumdepth \else
                      \protect\numberline{}{\csname the#1\endcsname}\fi
                      . #7}}\fi
          \@xsect{#5}}
\makeatother%
%%%%%%%%%%%%%%%%%%%%%%%%%%%%%%%%%%%%%%%%%%%%%%%%%%%%%%%%%%%%%%%%%

\appendix
\label{app-sec}
\section{The symmetric matrix $G(u_t)=e^{-Au_t}$}
\label{Green-mat}
The symmetric matrix $G(u_t)$ is given by
\begin{equation}
G(u_t)=\begin{pmatrix}
G_{11}(u_t) &&G_{12}(u_t)\\
G_{12}(u_t)&&G_{22}(u_t) 
\end{pmatrix},
\end{equation}
in which
\begin{align*}
G_{11}(u_t)&=-\dfrac{1}{4r_2}[1-2r_2-(1+2r_2)e^{2r_2u_t}]e^{-\lambda_1u_t},\\
G_{12}(u_t)&=-\dfrac{1}{2r_2}\delta(1-e^{2r_2u_t})e^{-\lambda_1u_t},\\
G_{22}(u_t)&=\dfrac{1}{4r_2}[1+2r_2-(1-2r_2)e^{2r_2u_t}]e^{-\lambda_1u_t},
\end{align*}
where $u_t=t/\tau_\gamma$, $r_1=\frac{1+2\delta}{2}$, $r_2=\frac{\sqrt{1+4 \delta^2}}{2}$, $\lambda_1=r_1+r_2$, and $\lambda_2=r_1-r_2$. Clearly, $\lambda_1>\lambda_2>0$ for all $\delta>0$.

\section{Mean of $U(t)$}
\label{mean-U-sec}
The mean of $U(t)$ is given in \eref{mean-U}. After some calculation, one gets
\begin{align*}
\overline{\langle x_A\rangle}=&\dfrac{F\tau_\gamma}{8 \gamma \delta r_2}e^{-2 r_1 u_t}\bigg[4r_2e^{2 r_1 u_t}[2-\theta+2\delta+\delta\theta(u_t-2)]+e^{\lambda_2 u_t}[(1-2r_2)(2-\theta)\\&+(1-\theta)\{4\delta^2+2\delta(1-2r_2)\}]-e^{\lambda_1 u_t}[(1+2r_2)(2-\theta)+(1-\theta)\{4\delta^2+2\delta(1+2r_2)\}]\bigg],\\
\overline{\langle x_B\rangle}=&\dfrac{F\tau_\gamma}{4 \gamma r_2}e^{-\lambda_1 u_t}\bigg[1+2(\delta-r_2)(1-\theta)-e^{2 r_2 u_t}[1+2(1-\theta)(\delta+r_2)]+2 r_2 e^{\lambda_1 u_t}[2+\theta(u_t-2)]\bigg],
\end{align*}
where the dimensionless parameter $\theta=2v\gamma/F.$

\section{Mean and correlation of $W_A$ and $W_B$}
\label{mean-corr}
The mean and correlation at time $\tau$ given in \erefss{mean-A}--\erefs{corr-AB} can be obtained as
\begin{align*}
\mu_A=&\dfrac{\alpha^2\tau_\gamma}{8\delta r_2}\bigg[4r_2[2(1+\delta)+\theta(\delta u_\tau-1-2\delta)]+[\{2-\theta+2\delta(1-\theta)\}(1-2r_2)+4\delta^2(1-\theta)]e^{-\lambda_1u_\tau}\\&-[\{2-\theta+2\delta(1-\theta)\}(1+2r_2)+4\delta^2(1-\theta)]e^{-\lambda_2u_\tau}\bigg],\\
\mu_B=&-\dfrac{\alpha^2\theta\tau_\gamma}{2}\bigg[(1-\theta)u_\tau+[1-2 r_2+2\delta(1-\theta)+2\theta r_2]\dfrac{(1-e^{-\lambda_1u_\tau})}{4r_2\lambda_1}\\&-[1+2 r_2+2\delta(1-\theta)-2\theta r_2]\dfrac{(1-e^{-\lambda_2u_\tau})}{4r_2\lambda_2}\bigg],\\
C_{AA}=&\dfrac{\alpha^2\tau_\gamma}{2\delta r_2}\bigg[4(1+\delta)r_2+[(1+\delta)(1-2r_2)+2\delta^2]e^{-\lambda_1 u_\tau}-[(1+\delta)(1+2r_2)+2\delta^2]e^{-\lambda_2 u_\tau}\bigg],\\
C_{BB}=&\dfrac{\alpha^2\theta^2\tau_\gamma}{16\delta r_2^2}\bigg[8r_2^2[2\delta(u_\tau-2)-1]+[1-2r_2+4\delta(1-r_2)+4\delta^2(1+4\delta-4r_2)]e^{-\lambda_1 u_\tau}\\&+[1+2r_2+4\delta(1+r_2)+4\delta^2(1+4\delta+4r_2)]e^{-\lambda_2 u_\tau}\bigg],\\
C_{AB}=&0,
\end{align*}
where $u_\tau=\tau/\tau_\gamma$, and $\alpha=F/\sqrt{\gamma T}$.

\vskip 2cm

\bibliographystyle{unsrt}
\bibliography{ref.bib}
\end{document}